\documentclass[useAMS]{mn2e}
\usepackage{graphicx}
\usepackage{color}

\title[BCGs in the SDSS DR6]{The properties of Brightest Cluster Galaxies in the SDSS DR6
adaptive matched filter cluster catalogue}

\author[A.Pipino et al.]{A. Pipino\thanks{antonio.pipino@phys.ethz.ch}$^{1,2,3}$, T.Szabo$^3$ \& E. Pierpaoli$^3$, S.M.MacKenzie$^{2,4}$ \& F.Dong$^5$ \\
$^1$Institure for Astronomie, ETH Zurich, 8093 Zurich, CH \\
$^2$Department of Physics and Astronomy, University of California Los Angeles, Los Angeles CA 90025, USA\\
$^3$Department of Physics \& Astronomy, University of Southern California, Los Angeles 90089-0740, USA\\
$^4$Department of Physics and Astronomy, University of Louisville, Louisville KY 40292, USA\\
$^5$Department of Astrophysical Sciences, Princeton University, Princeton, NJ 08544, USA}

\long\def\symbolfootnote[#1]#2{\begingroup%
\def\thefootnote{\fnsymbol{footnote}}\footnote[#1]{#2}\endgroup}

\begin{document}

\date{}

\pagerange{\pageref{firstpage}--\pageref{lastpage}} \pubyear{2008}

\maketitle

\label{firstpage}

\begin{abstract}
We study the properties of Brightest Cluster Galaxies (BCGs) 
drawn from a catalogue of more than 69000 clusters in the SDSS DR6 
based on the adaptive matched filter technique (AMF, Szabo et al., 2010).
Our sample consists of more than 14300 galaxies in the redshift range 0.1-0.3.
We test the catalog by showing that it includes well-known BCGs which
lie in the SDSS footprint.
We characterize the BCGs in terms of r-band luminosities and optical colours as well as their trends
with redshift. In particular, we define and study the fraction of \emph{blue} BCGs,
namely those that are likely to be missed by either colour-based cluster surveys
and catalogues, as shown by a direct comparison to maxBCG clusters that are
matched in the Szabo et al. catalogue.
We further compare the properties of the BCGs to those of the second
and third brightest galaxies in the same cluster.
Finally, we morphologically classify those galaxies hosted in the richest clusters.

We find that the BCG luminosity distribution is close to a Gaussian,
whose mean has a redshift evolution
broadly consistent with pure aging of the galaxies. Richer clusters
tend to have brighter BCGs, however less \emph{dominant} than in poorer
systems.
4-9\% of our BCGs are at least 0.3 mag bluer
in the g-r colour than the red-sequence at their given redshift.
Such a fraction decreases  to 1-6\% for clusters above a richness of 50,
where  3\% of the BCGs are 0.5 mag below the red-sequence. 
A preliminary morphological study suggests that the increase
in the blue fraction at lower richnesses may have a non-negligible contribution
from spiral galaxies. 
%Therefore we suggest a cut at richness
%of 100 (and $M_r < -22.5$ mag) if one wants to focus only on early-type galaxies. 
In terms of redshift evolution, the overall blue fraction goes from $\sim$5\% in the redshift range 0.1-0.2 to $\sim$10\% in the redshift bin 0.2-0.3. 
The blue fraction seems to increase at higher redshifts, however the scatter in the colours
and the fact that the catalog is no longer complete hamper us from having firm conclusions.
We show that a colour selection based on the g-r red-sequence or on a cut at colour u-r$>2.2$ can lead to missing the majority of such blue BCGs.
Finally, the blue fraction increase by a factor 1.5 at most when the study is extended
to the three brightest galaxies of each cluster.

We also extend the colour analysis to the UV range by cross-matching
our catalogue with publicly available data from Galex GR4 and GR5.
We show a clear correlation between offset from the 
optical red-sequence and the amount of UV-excess.

Finally, we cross-matched our catalogue with the ACCEPT cluster sample (Cavagnolo et al., 2009), and
find that blue BCGs tend to be 
in clusters with low entropy and short cooling times. %and typically with
%a $L_X$ above the mean for their temperature. 
That is, the blue light
is presumably due to recent star formation associated to gas feeding by cooling flows.
\end{abstract}


\section{Introduction}

The Brightest Cluster Galaxies (BCGs) are the most massive galaxies in the universe,
with most of their stellar mass in place by redshift 2.
Therefore, they are expected to experience the galaxy formation process in the most extreme way. Namely they should form at earlier times, more rapidly and with a more intense 
star formation event than lower mass cluster members.
While such a formation scenario is naturally explained
in the framework of the revised \emph{monolithic collapse models} (e.g. Larson, 1974, Pipino et al., 2008), it seems more difficult to reconcile
their existence within the \emph{hierarchical growth scenario} where the largest structures
are the last to form.
A possible way out is that 
feedback halted the star formation at very early times and the mass assembly of BCGs can be 
simply explained with a series of gas-less mergers of old stellar systems (De Lucia \& Blaizot, 2007, but see Whiley et al., 2008, Pipino et al., 2009b, Pipino \& Matteucci, 2008).
A closer look at their stellar populations tells us that BCGs
have similar mean stellar ages 
and metallicities to non-BCGs ellipticals of the same mass but they have somewhat higher $\alpha$/Fe ratios, indicating that star formation may have occurred over a shorter time-scale in the BCGs
(von der Linden et al., 2007, Loubser et al. 2008). Moreover, 
they depart from the Faber-Jackson relation for ellipticals (Faber \& Jackson, 1976) and seem to be larger and with a higher stellar velocity dispersion
than ellipticals of the same mass (von der Linden et al., 2007, 
Bernardi et al., 2007, 2008). Furthermore, the 
BCG luminosity function differs from the usual Schechter (1976) form that holds for normal
cluster members (e.g. Hansen et al., 2005), in that it can be modelled as a Gaussian whose mean increases with the cluster richness (Lin et al., 2004).
Therefore, a study of the colours in BCGs as opposed to ``ordinary'' early-type
galaxies out to high redshift 
is a test bench to discriminate among models for galaxy formation (e.g. Roche et al., 2009). 
In particular, the colour evolution may place constraints
to the time and the intensity of their last star formation episode, whereas a joint
analysis of the colour evolution with the thermal status of the surrounding intracluster
medium may tell us what halted the gas supply and inhibited further star formation.

Luminous red galaxies - as most of the BCGs are - are used as a good tracer of the underlying dark matter distribution (Ho et al., 2009a, Reid \& Spergel, 2009). Their properties are strongly linked to the host halo mass, so their census can 
provide us with a map of large-scale over-densities in the Universe. They have been used
to detect baryonic acoustic oscillations (e.g. Sanchez et al, 2009, and references therein),
and it has been put forward (Ho et al., 2009b) that a combination of a galaxy redshift survey such as SDSS and a CMB survey
can be used as a method for detecting the missing baryons.
Therefore BCGs can be a promising tool for precision cosmology as well.

Finally, since BCGs occupy a special place in that they sit at the bottom of the cluster potential well,
we expect their present-day properties to be linked to the state of the intracluster gas. Recent studies have reported
examples of ongoing star formation in the massive central galaxies
of cool core clusters (Crawford et al. 1999, Edge 2001, Goto 2005,
McNamara et al. 2006, Hicks \& Mushotzky 2005, O'Dea et al., 2006, Edwards et al., 2007, 2009).  Bildfell et al. (2008) found
that the presence of optical blue cores in 25\% of its BCG sample is directly linked to the
X-ray excess of the host clusters. Moreover the position of these BCGs
coincide with the peak in X-ray emission. 
Their interpretation is that the recent
star formation in BCGs is associated with the balance between heating and cooling in the
ICM in the sense that the clusters that are actively cooling are forming
stars in their BCGs. Other evidence comes from Rafferty et al (2008),
Cavagnolo et al. (2008). In particular, Pipino et al. (2009a)
demonstrated a one-to-one correspondence between
blue cores in BCGs and a UV-enhancement observed
using GALEX.
The blue light coming
from the cores might render the BCGs 0.5 - 1 mag bluer than
the \emph{g-r} red-sequence (Bower et al., 1992, Baldry et al, 2004). This may impact the creation of large
optical cluster catalogue based on the presence of a well
defined red-sequence as well as on the presence of a BCG.

Szabo et al. (2010) have created the largest available catalogue of optically
selected clusters from the SDSS DR6. This cluster catalogue
is based on the matched filter method (Dong et al., 2008) and
does not include any colour selection of member galaxies.
An interesting by-product of the cluster finder algorithm is the
creation of the largest available catalogue of Brightest Cluster
Galaxies with homogeneous photometry (and photometric redshift)
without any selection on colours.
In the following we will refer to the Szabo et al. (2010)
cluster catalogue and the related BCG catalogue as synonyms, since
each BCG is uniquely associated to a cluster.

The main aim of this paper is to characterize such a catalogue by 
describing some tests done to check the its accuracy
and characterizing the BCG properties in the colour-colour
and colour-magnitude spaces.
By means of a redshift-independent definition of \emph{blue} BCGs, we will
be able to quantify the bias that affects catalogues created via a colour-based selection.

Moreover, with such a BCG catalogue we are in the position of pursuing several main goals.
For instance, in this paper we extend the catalogue by including UV-optical colours
of the BCGs thanks to Galex public data. 
%Also, the fact that we do not exclude blue BCGs and that these galaxies tend
%to lie very close to the X-ray cluster centres, enhance the
%cross-correlation and 
By means of a positional cross-matching of our catalogue with
published compilations of X-ray selected clusters 
%with respect to other works. Therefore 
we are in the position of 
extending Bildfell et al.(2008)'s and Pipino et al.(2009a)'s results to a larger sample
of clusters/BCGs.

A quantification of how cooling flows and blue BCGs impact cluster detection
from Sunyaev-Zeldovich surveys is addressed in a companion paper (Pipino \& Pierpaoli, 2010).

The scheme of the paper is thus the following: in Sec.~\ref{cat} we briefly describe the Szabo et al. cluster catalogue,
briefly summarizing the BCGs selection and properties. 
In Secs.~\ref{char} 4 and 5
we will present the characteristics of the BCG sample in terms of luminosity functions, colours
and redshift evolution of these properties as well
as we compare them to other existing catalogues. We apply the catalogue to the study of the UV-optical
colours and the X-ray properties of our BCGs in Sec.~\ref{uso}. Conclusions will be drawn in
Sec.~\ref{concl}.

\section{The catalogue}
\label{cat}

In this section we briefly summarize how the Szabo et al. cluster catalog
that we use has been built. A detailed description
can be found elsewhere (Szabo et al., 2010).

\subsection{Data}

The data on which Szabo et al. catalogue has been built are from SDSS DR6
(Adelman-McCarthy et al., 2008).
All galaxy measurements were extracted from the
{\it Galaxy} view on the CasJobs DR6 database\footnote{http://cas.sdss.org/dr6}.
Szabo et al. selected only galaxies that are detected in 1x1 binned
images, that have a measurable profile, that are not saturated
or contain peaks other than that provided by the estimator
of the SDSS pipeline.

The adopted photometric redshifts are based on a neural network \emph{cc2} estimator and are
available in the table {\it Photoz2}  (Oyaizu et al., 2008). We made this choice
because of their more reliable error estimates and lack of evident colour biases  
as opposed to those available from {\it Photoz} (Csabai et al., 2008). For more details
we refer to Szabo et al. (2010)'s paper.
Absolute magnitudes are calculated
by means of the {\it kcorrect} code v4.1.4  (Blanton \& Roweis, 2007).
In the following we will always use SDSS \emph{model} magnitudes.

\subsection{The cluster catalogue}
\label{met}

\begin{figure}
%\epsscale{.80}
\includegraphics[width=8cm,height=4cm]{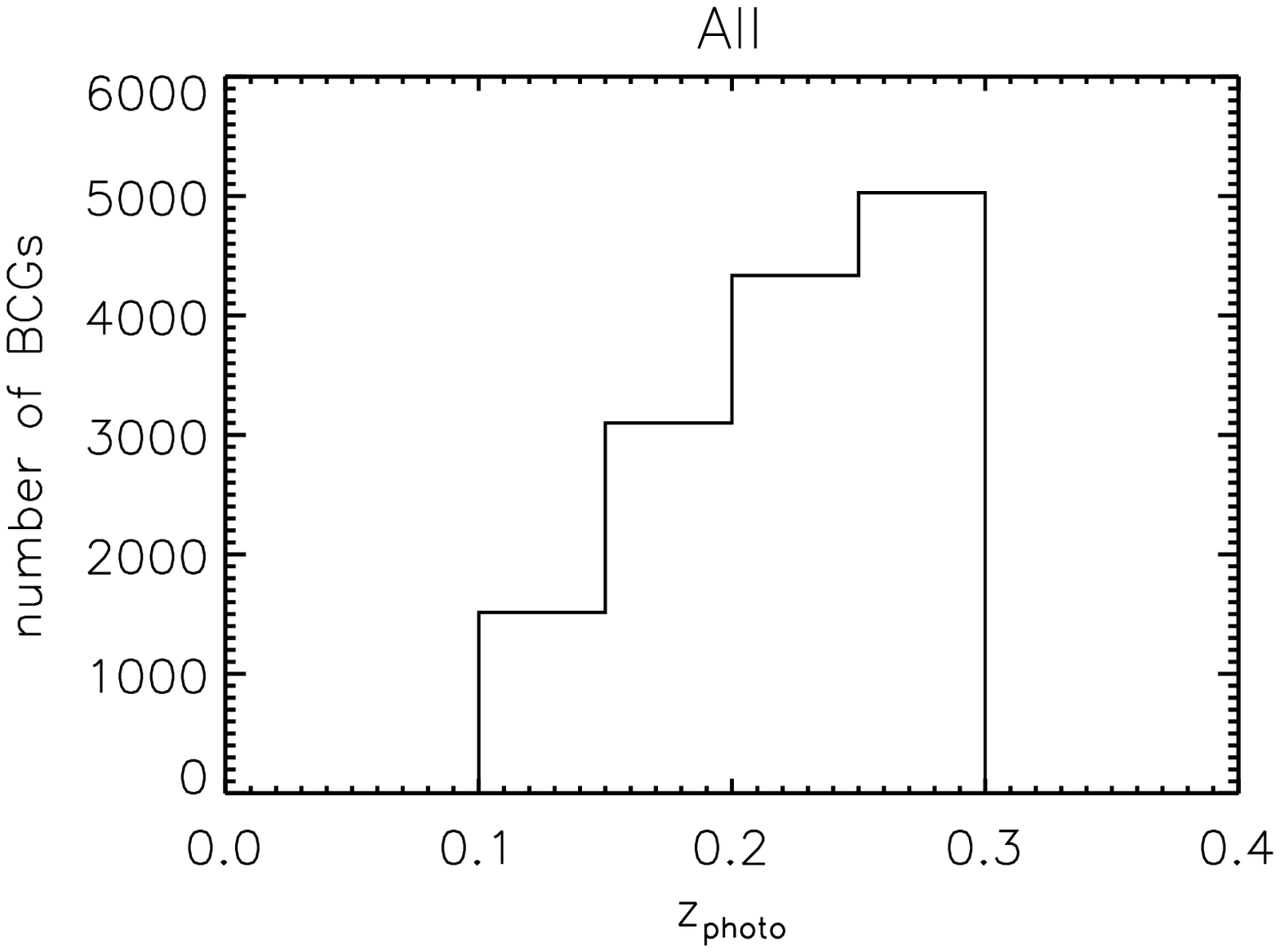}
\includegraphics[width=8cm,height=4cm]{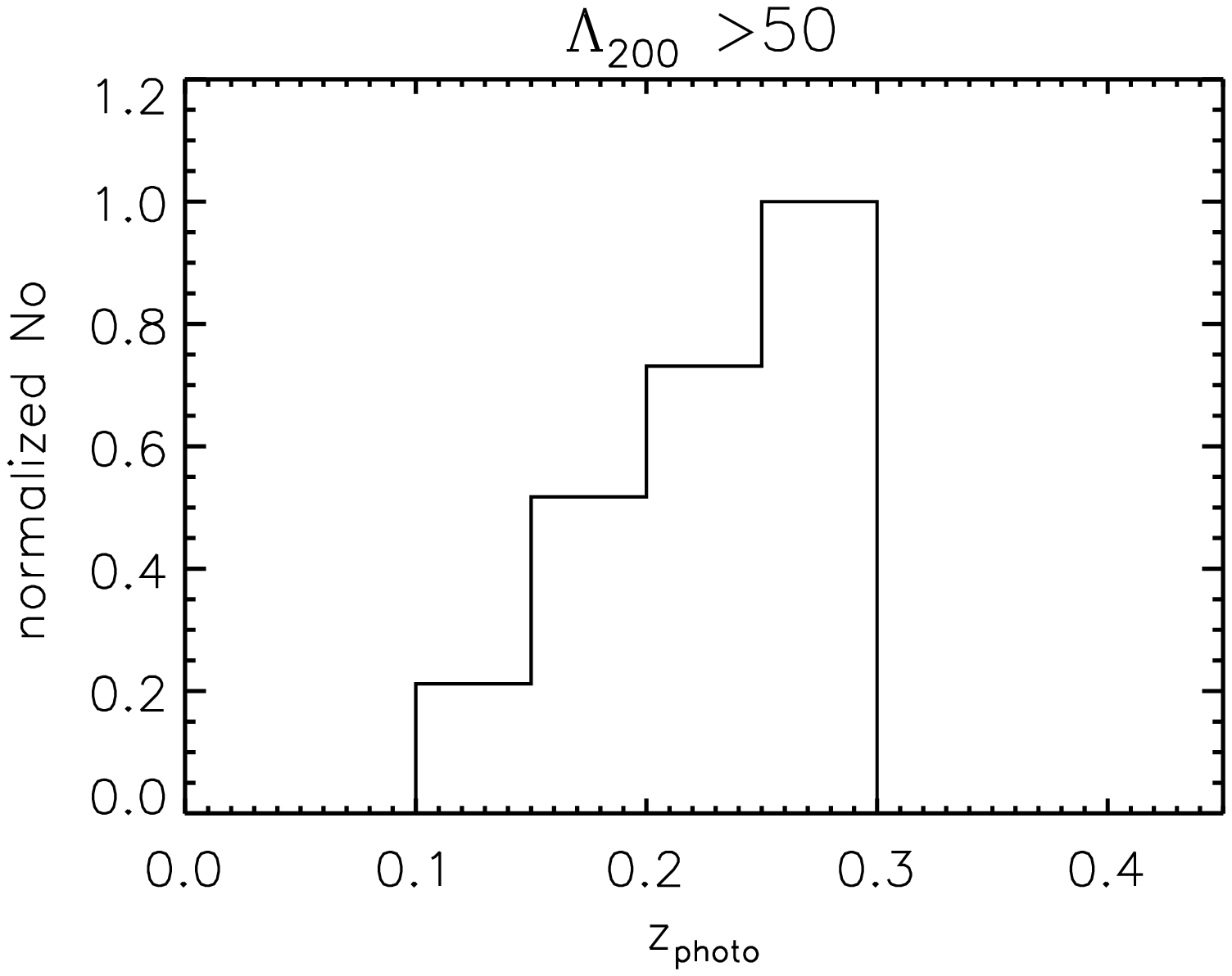}
\caption{Redshift distribution of all the BCGs and those in rich clusters.}
\label{fig6}
\end{figure}

The matched filter method (Kepner et al., 1999) that Szabo et al. (2010) use is presented
in detail by Dong et al (2008). In practice, it is a likelihood method which identifies clusters 
by convolving the optical galaxy survey with a set of filters based on a modeling of the cluster 
and field galaxy distributions. A cluster radial surface density profile, a galaxy luminosity 
function, and redshift information (when available) are used to construct filters in position, 
magnitude, and redshift space, from which a cluster likelihood map is generated. The peaks in 
the map thus correspond to candidate cluster centers where the matches between the survey data 
and the cluster filters are optimized. The algorithm automatically provides the probability for 
the detection, best-fit estimates of cluster properties including redshift, radius and richness, 
as well as membership assessment for each galaxy. 
Usage of the apparent magnitudes and the redshift estimates instead of simply 
searching for projected galaxy over-densities suppresses the foreground-background 
contamination.
 
The cluster catalog is constructed with an iterative procedure.
The process starts from a density model of a smooth background with no clusters. For each galaxy 
position, we then evaluate the likelihood increment we would obtain by assuming that there is in fact 
a cluster centered on that galaxy. At each iteration, the cluster candidate 
which resulted in the greatest likelihood increase is retained. 
A list of cluster candidates then becomes available in decreasing order of detection likelihoods.
%The search radius for galaxies belonging to the cluster is set to 
%be the virial radius of the cluster, or more specifically here, 
The cluster richness $\Lambda_{200}$ is then 
defined to be the total luminosity in units of $L^*$ inside $r_{200}$, namely the radius inside which the mass over-density is 200 times 
the critical density. %, i.e., 200$\Omega_M^{-1}$ times the average background. % \citep{Evr02}.

For the magnitude filter, Szabo et al adopt a luminosity profile described by a central galaxy plus
a standard Schechter luminosity function (Schechter, 1976).

Dong et al. (2008) showed that the selected cluster sample is 
$\sim 85\%$ complete and over $90\%$ pure for systems more massive than 
$1.0\times10^{14} h^{-1}$ M$_\odot$ ($\Lambda_{200} \sim 50$) with redshifts in the range 0.1-0.4. 
In order to have a reliable assessment of the BCG colours, we
restrict the sample to galaxies in the redshift range [0.1,0.3],
where the intrinsic scatter in the observed g-r colour is the smallest.
The estimated 
cluster redshifts derived from maximum likelihood analysis show small errors 
with $\Delta z < 0.01$.

%\subsection{The cluster catalogue}

The final Szabo et al.'s catalogues has more than 69000 entries.
We refer the reader to the main catalogue paper for details on the cluster
characteristics and comparison with other automated catalogues.
A complete version of the catalogue with the three brightest galaxies
from which we derive the sample discussed in this paper can be found Szabo
et al. (2010).
%TO REINSTATE WHEN/IF THE WEBSITE IS READY:
%or at the following URL: http://aran.usc.edu/~XXXX.html 

\subsection{A test of the accuracy of the catalogues}
\label{test}

We tested the accuracy of our selection by cross-matching the position
of such an extended BCG sample with known coordinates of well studied BCGs taken from the literature (Crawford et al., 1999,
Bildfell et al., 2008, Loubser et al., 2008 as well as the SIMBAD database) whose position is within the region of sky
covered by SDSS DR6 and that lie at $0.1<z<0.3$.
While we will mostly focus on BCG as the ``brightest'' member in every group
or cluster of galaxies, it is important to 
consider also the 2nd and 3rd most luminous galaxies in a cluster (by using the r-band magnitude luminosities) in the following exercise. This is needed because the above mentioned works
either defined the BCG as the brightest galaxy in a band other than the \emph{r} one or
by their position in the cluster. Therefore there are a few cases in which
their BCG is either the second or the third brightest in our definition.

Our sample matches 73\% of the BCGs in the Crawford et al. (1999) list with an accuracy of less than 0.3'' in angular position and less of 0.01 in redshift space,
and more than 80\% with if we require an accuracy of a few arc seconds, which is the typical error in the coordinates.
Also differences in coordinates among authors (and the SIMBAD database) 
for the same galaxy for the same galaxy may amount to a few arc seconds. 

In the remaining 20\% of the cases:\\
i) the known BCG that we want to match lies too close to the edge of the
sky region covered by the SDSS DR6 and the AMF finder has problems in identifying a cluster there.\\
ii) it is very close to the limit redshift range considered in this work. As an example
we mention the case when the BCG to be matched has redshift 0.1, but its host cluster has a redshift $<$ 0.1 and falls in the region
where our catalogue is not complete (and thus missed).\\
iii) it is part of a sub-structure of a bigger cluster that it is not \emph{resolved} by our cluster finder.\\
A further complication is that, in the above mentioned literature compilations, the authors often \emph{decide} which
galaxy to label as the BCG when galaxies with very similar luminosity were present.
In some cases their selection has been done on the basis of the presence
of an extended stellar halo which clearly made the BCG candidate a cD galaxy.
In other cases, the galaxy closer to the X-ray centre of the cluster has been
selected. Finally, there are cases in which a galaxy cluster is made of
sub-clusters in the act of merging, each one with its own BCG.
Therefore it is not surprising that we have a fraction of cases in which the positional best match with
a known BCG has been obtained
by the second or the third most luminous galaxies in the entire Szabo et al.'s BCG sample.
Also, we have cases in which the first and the second brightest galaxies of a given cluster
in our entire BCG sample match the BCGs of two known (merging) sub-clusters.
A more detailed assessment of these cases will be the topic of a forthcoming study.

To our knowledge, the Szabo et al. cluster catalogue is the only one tested
in comparison to known cluster \emph{and} BCG positions.
Other works (e.g. Koester et al., 2007)
%return a similar success rate. However these works typically 
limited their analysis to matching the position of their clusters
with the centre of a known cluster. {Here we show that our catalog includes the single galaxies
classified as BCGs from previous works.}
%We also note that great care has been devoted to verify that known extremely blue BCGs
%(such as the one in Abell 1835 and other cases in the Bildfell et al., 2008, sample) are present in our catalogue.

\subsection{The final BCG sample}
\label{select}

We will focus on BCG as the brightest\footnote{as opposed to other
selection criteria such as being the highest likelihood member or the closest to the
estimated cluster centre.} member in every group
or cluster of galaxies in the r band. 
We further exclude from the analysis galaxies whose error on the g-r colour, expressed as $3\cdot \sqrt{\sigma_g^2+\sigma_r^2}$,
exceeds 0.3 mag. They represent about less than  1\% of the galaxies in the redshift range 0.1-0.3.
The final sample comprises of 14344 galaxies.
%However, \emph{true} BCGs are preferentially found in rich clusters and
%have a luminosity larger than $\sim L^*$. 
%We will refer to the brightest among the three 
%as either \emph{true} BCGs (\emph{tBCGs}) or \emph{1st ranked} galaxies. We have 15242 of such galaxies in clusters with richness
%above 30 and 4556 in clusters with richness above 50. They have absolute magnitudes well above $M^*-2$.

The number of BCGs (Fig.~\ref{fig6}) increases with redshift as the underlying
cluster redshift distribution. 
The richness of the cluster does not play a role: the histogram for rich clusters
tracks the one for the whole population.

\begin{figure}
%\epsscale{.80}
\includegraphics[width=8cm,height=8cm]{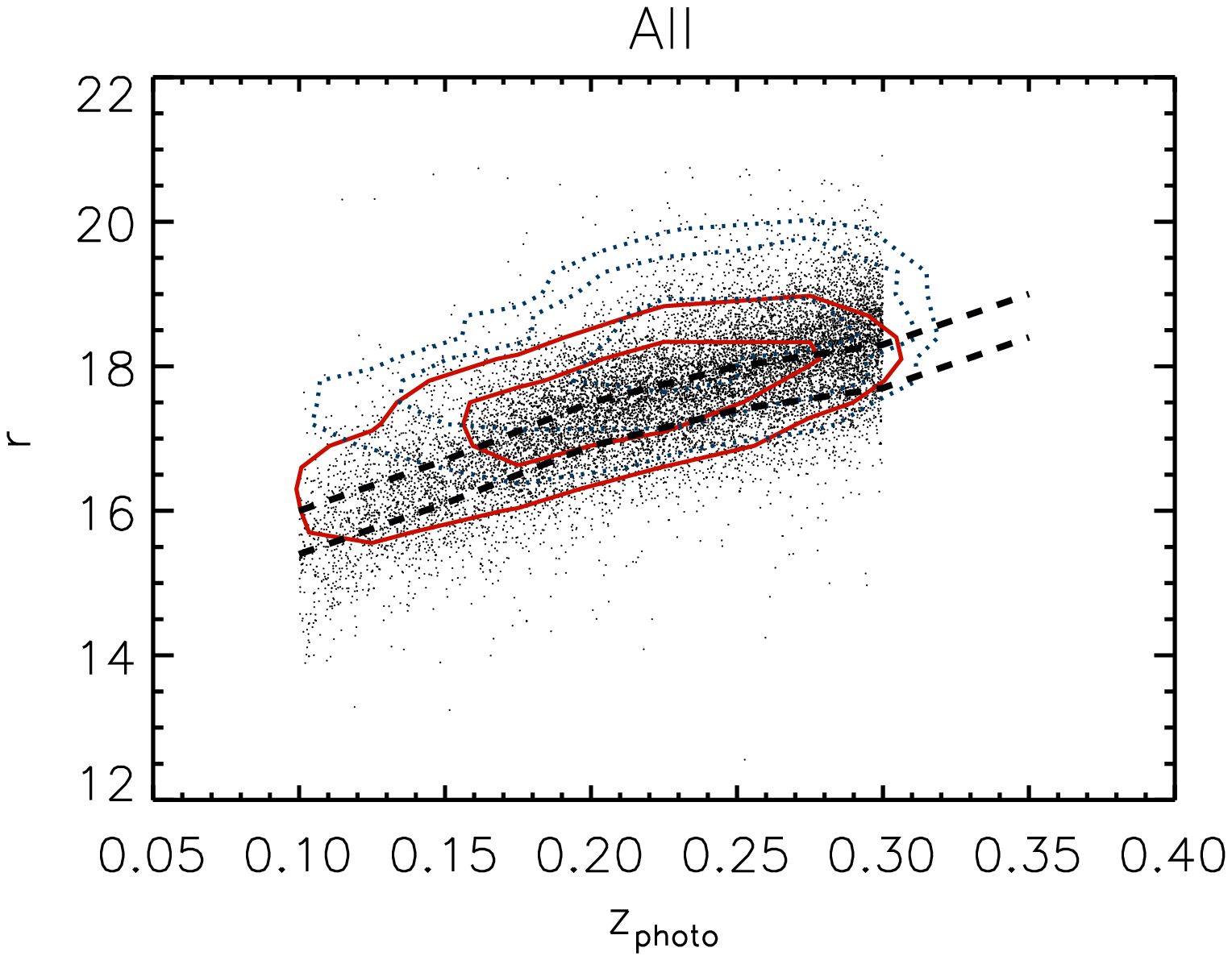}
\includegraphics[width=8cm,height=8cm]{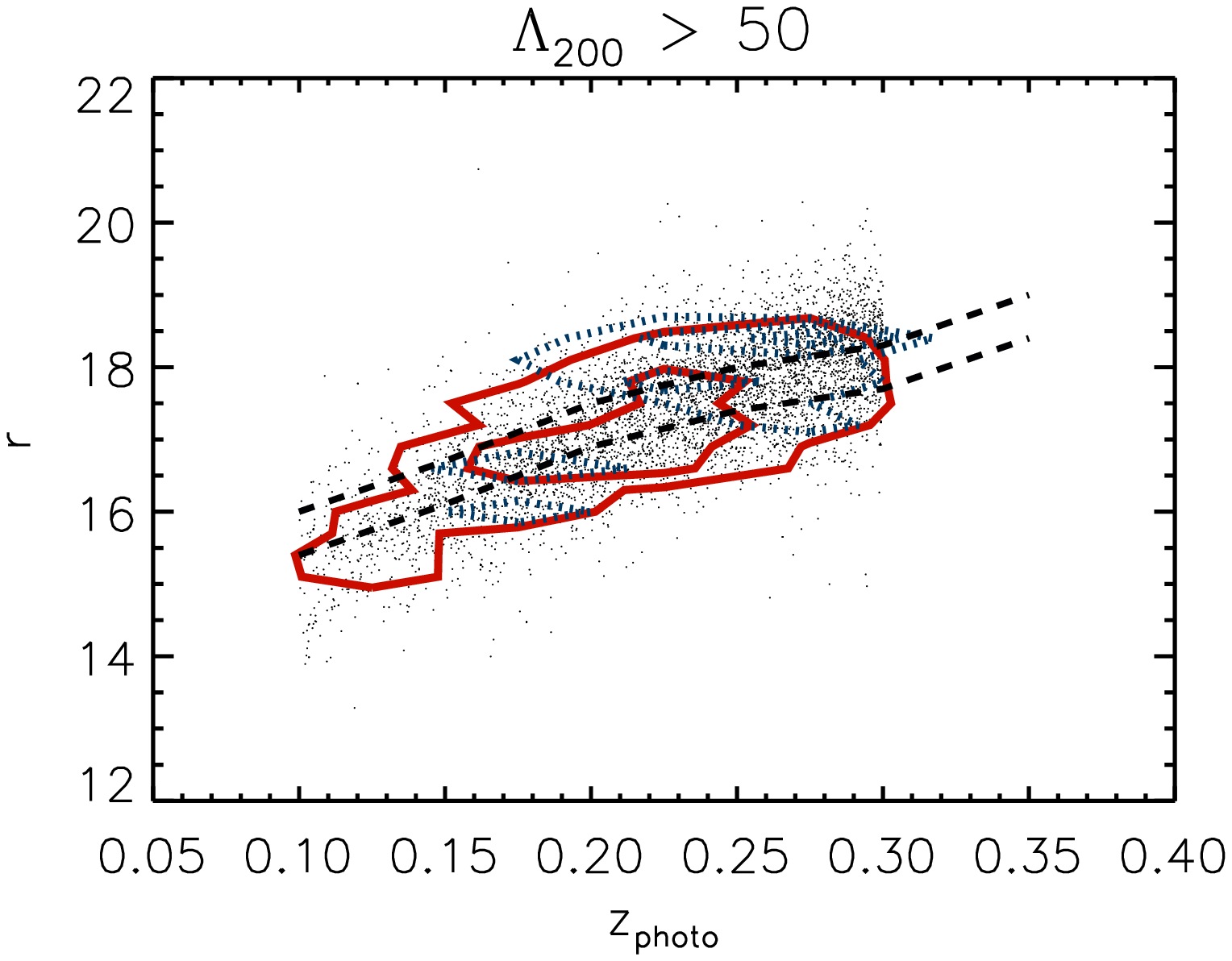}
\caption{Distribution of galaxies in the plane \emph{r}-band magnitude versus photometric redshift: all BCG (points and solid contours, upper panel)
and galaxies in clusters with richness larger than 50 (solid contours, lower panel).
BCGs bluer than 0.3 mag from the red-sequence at their redshift are shown by dotted iso-density (number of galaxy/bin area) contours. 
Dashed lines: 1$\sigma$ region around the mean relation by Loh \& Strauss (2006) for BCGs in SDSS. 
Compare with Koester et al (2007) Fig.3.}
\label{fig1}
\end{figure}

\section{BCG characterization}
\label{char}

In this section we will focus on the BCG characterization
in terms of luminosity, colours and explain how the inclusion of \emph{blue}
BCGs make Szabo et al. (2010) catalogue different from
others in the literature. For a comparison of BCG luminosity distributions
between catalogues we refer to Szabo et al. (2010).
From Sec.~\ref{col} we anticipate that we refer to galaxies that are 0.3 mag
below (i.e. bluer) than the g-r red-sequence at their respective redshifts as \emph{blue} BCGs.

\subsection{BCG luminosity}
\label{lf}

In Fig.~\ref{fig1}, we compare our
BCGs to the BCG magnitude-redshift relation inferred by Loh \& Strauss (2006) for LRG
in the redshift range 0.12-0.38.
The dashed lines bracket the 1$\sigma$ scatter around this relation.
At high richnesses (lower panel) the spread in our 
BCG sample is comparable with a $\sim$3$\sigma$
scatter around Loh \& Strauss' relation as well as the mean trends look very similar to each other.
In particular, we find that $r \sim 12.6 \cdot z + 14.7$.
A substantial population of blue BCGs is present when considering the entire sample (upper panel)
and offsets our distribution towards fainter magnitudes with respect to Loh \& Strauss (2006) findings.

\begin{figure}
%%\epsscale{.80}
\includegraphics[width=8cm,height=7cm]{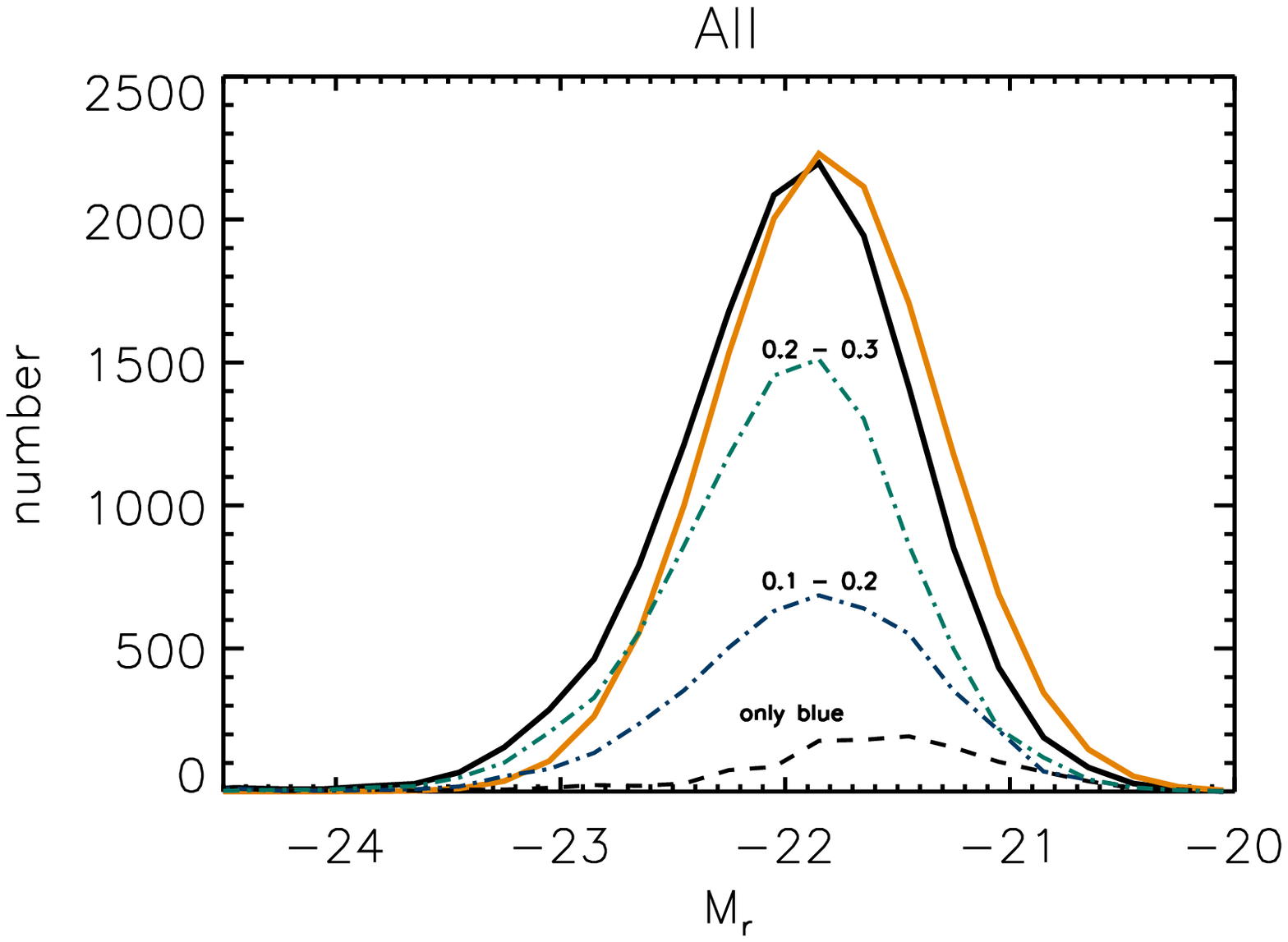}
\includegraphics[width=8cm,height=7cm]{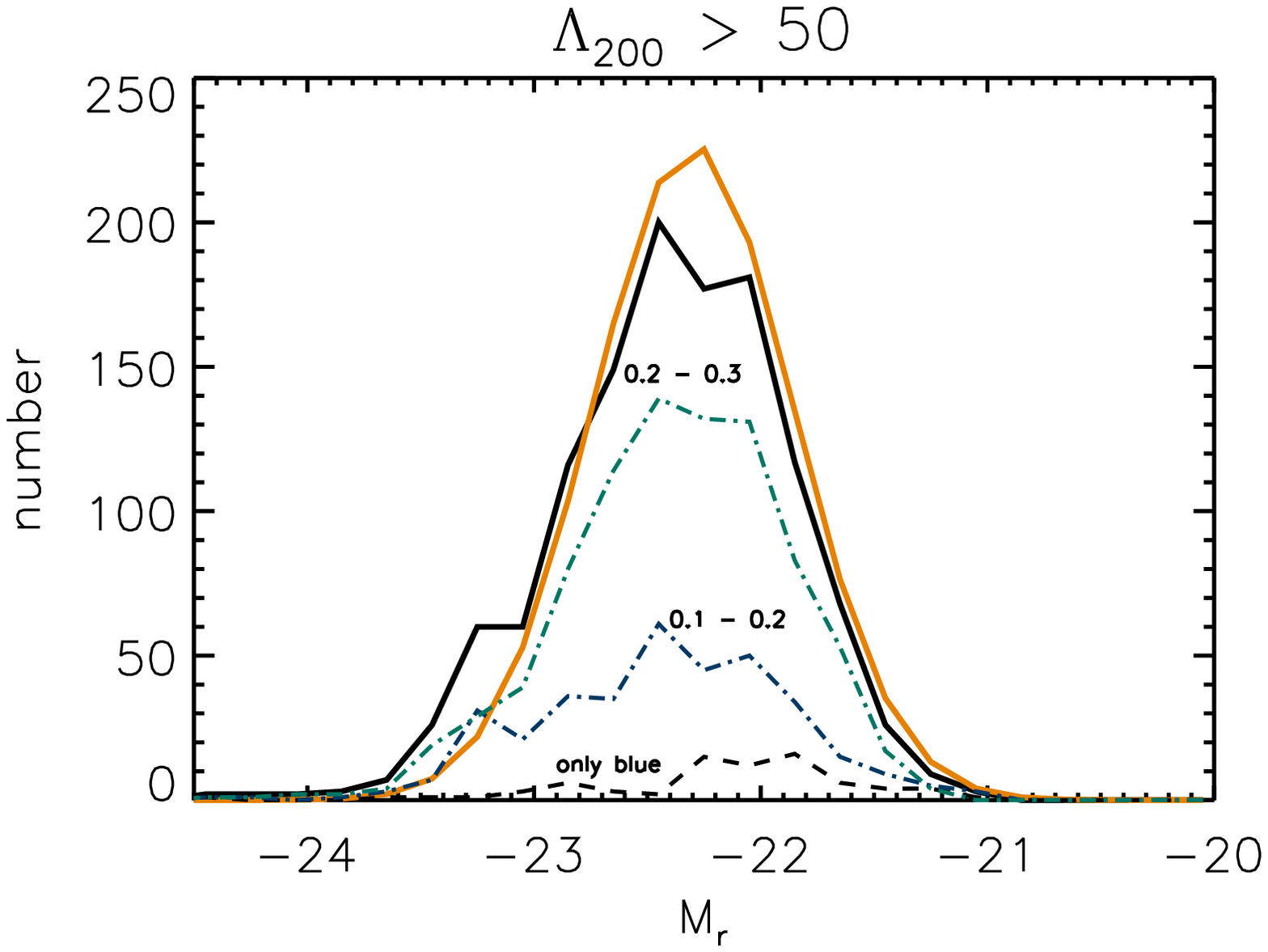}
\caption{\emph{Upper panel.} Solid thick line: luminosity distribution the r-band for BCGs in the redshift range
[0.1,0.3]. Lighter thick line: Gaussian distribution with mean equal to the average of the above distribution 
and $\sigma$=0.5 mag. The function for \emph{blue} BCGs (dashed line) and the subsets
of all BCGs divided into two redshift bins (dot-dashed lines)
are also shown. 
\emph{Lower panel}:  As above, but only for rich clusters ($\Lambda_{200} > 50$).}
\label{fig2}
\end{figure}

In Fig.~\ref{fig2} (upper panel) we show the distribution in luminosity in the r-band for the BCGs in the redshift range
[0.1,0.3]  as a thick solid line. This curve slightly deviates from a Gaussian distribution with mean equal to the average
$M_r$ in the same redshift range and $\sigma\sim$0.5 mag which is commonly adopted as the BCG luminosity function (e.g. Hansen et al., 2005).
%A significant contribution to the asymmetry at the faint side is due to the blue BCGs, whose
%distribution is shown by a dashed line. 
As we will discuss in more details later in the paper,
bluer BCG tend to populate poorer systems. Since the BCG luminosity scales
with the cluster richness (Lin \& Mohr, 2004, Hansen et al., 2009), blue BCGs (dashed line) tend to be fainter than the average.
Note that here we are showing the distribution
for the entire BCG sample, irrespective of the galaxy redshifts. In fact,
when we look at the distributions in two redshift bins for the whole sample (dash-dotted lines),
we notice that the distributions are narrower.

The average $M_r$ increases (conversely the luminosity decreases) at smaller redshift. 
In particular we find that the mean $M_r$ scales linearly with log (time) as
expected from pure passive evolution (e.g. Tinsley, 1980, Nelson et al., 2001).
The average luminosity drops by $\sim$20\% from redshift 0.3 to 0.1.
Also highlighted (lower panel) are the distribution functions in clusters with richness above
50. A comparison between the two panels makes evident
that the tail at faint magnitudes is due to BCGs in low richness clusters.
While a study of the BCG luminosity function clearly deserves more attention, 
we can conclude that we found a distribution that it is broadly consistent with a Gaussian shape
and featuring a redshift evolution in agreement with a passive evolution of the BCG population (e.g. Brough et al., 2002).
A more quantitative assessment of the BCG luminosity function and its evolution would
require a more careful treatment of the data.
Here we rely on the absolute magnitude in the r-band calculated by means of the template
fitting approach (Blanton \& Roweis, 2007, Csabai et al. 2008) but with photometric
redshift derived from a neural network estimator (Oyaizu et al., 2008), namely not
is self-consistent way. Furthermore no evolutionary corrections have been applied.
Moreover, it is known that sky subtraction errors in the SDSS pipeline may significantly affect
the magnitude of the brightest objects (Adelman-McCarthy et al. 2008, and references therein).
Finally the catalogue is not complete at the poor richness end, and this 
implies that the low-luminosity tail of the luminosity distribution is not
correctly represented.
Finally, the effect of small number statistics is clear in the distribution
for rich ($\Lambda_{200} > 50$) cluster.

In concluding the section, we add that Fig.~\ref{fig2} suggests a cut
at $M_r < - 22.5$ if one wants to use a sample almost made by red BCGs.

\subsection{Colour classification}
\label{col}

In order to understand the difference between a cluster (and BCG) catalogue
which relies on the matched filter method with respect to a colour classification,
it is important to define and characterize the fraction of \emph{blue} BCGs.
In this work, galaxies 0.3 mag
below (i.e. bluer) than the average g-r colour (observed frame) for
early types at their respective redshifts are considered as blue BCGs. 
Since these galaxies obey to the color-magnitude sequence, and since
we focus on a narrow range in luminosities (at the high mass/luminosity-end),
in practice our requirement implies that these blue BCGs will
lie below the red-sequence at their given redshift.
Indeed, we chose such a cutoff since 0.3 mag is more than 6$\sigma$ off the mean value of the color-magnitude relation (Bower et al., 1992).
In practice, we self-consistently evaluate the average
g-r at a given redshift by means of our BCG sample and
we use it as a zero point for measuring the offset
(i.e. the blueness) of the single galaxies. Not surprisingly, it turns out that this coincides
with the mean trend of the g-r versus redshift curve reported by Blanton \&
Roweis (2007, c.f. their Fig.3), hence basically with the locus where galaxies in the Cut I sub-sample of the Luminous Red Galaxy (LRG) sample (Eisenstein
et al., 2001) lie. In other words, it is very likely that LRGs that are also BCGs in their own cluster, end up
as BCGs in the Szabo et al. catalogue. The converse is not true, since the colour cuts in the LRG sample get rid of the galaxies that, albeit
very bright, are 0.3 mag bluer than the average g-r colour.
The fraction of the galaxies bluer than the average with respect to the total number of BCGs gives
us an estimate of the colour bias - possibly introduced by recent star formation (and cooling flows) -
into red-sequence based cluster catalogues.

\begin{figure}
%\epsscale{.80}
\includegraphics[width=9cm,height=6cm]{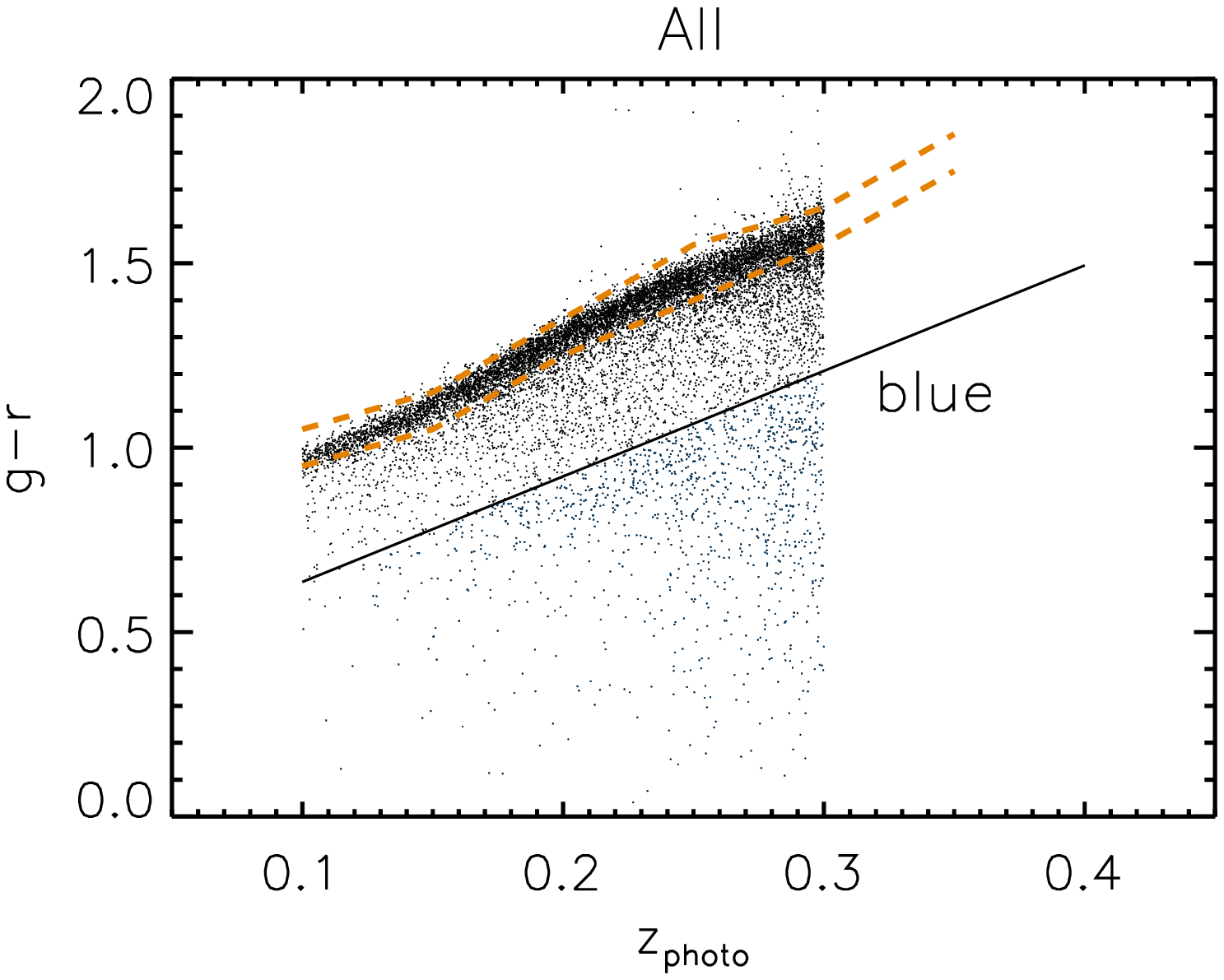}
\includegraphics[width=9cm,height=6cm]{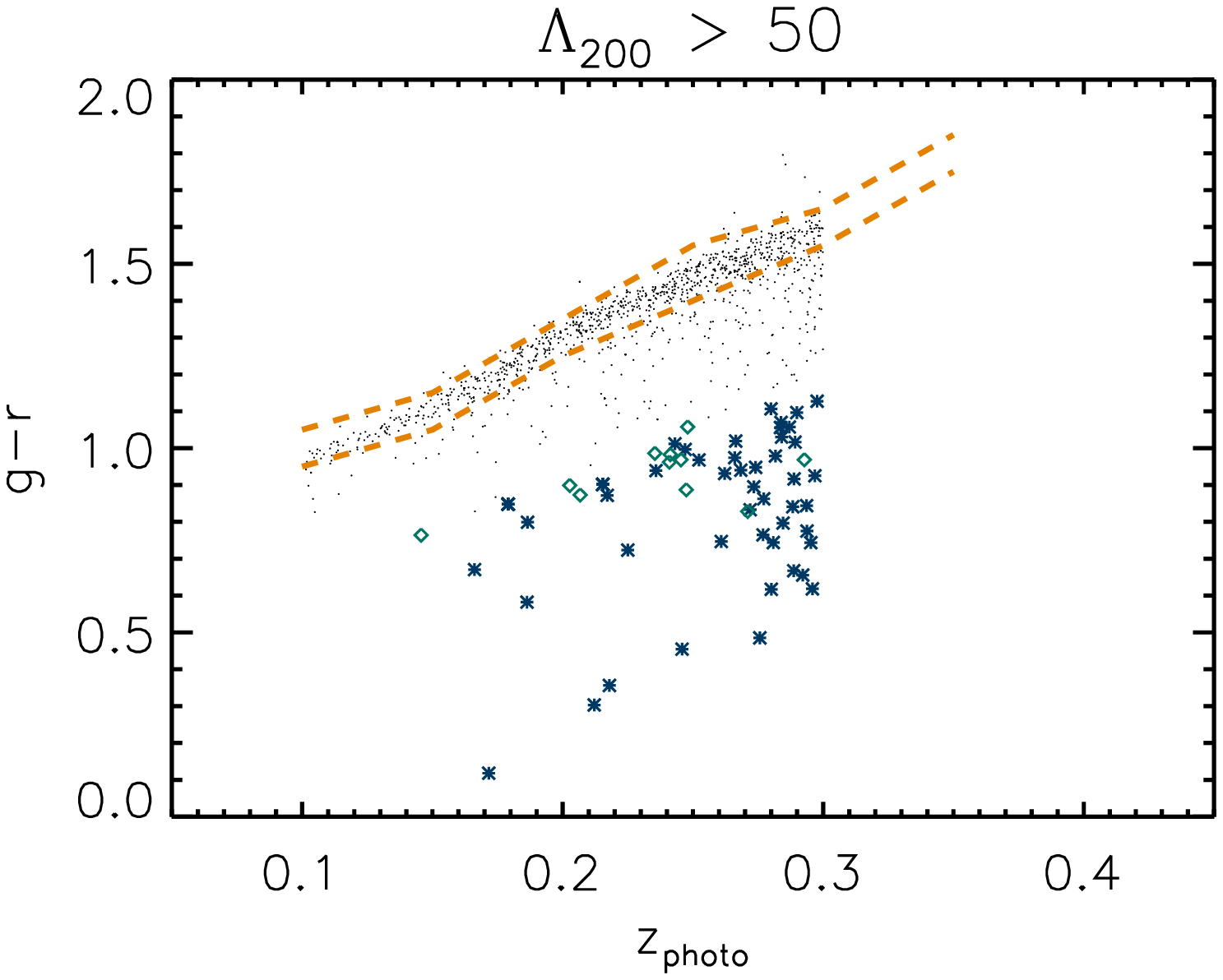}
\caption{\emph{g-r} color as a function of photometric redshift for both all BCGs 
and those in rich clusters. 
%The bottom-left and the top-right panels refer to different
%cuts in the cluster richness. Bottom-right panel: BCGs in clusters in common with maxBCG catalogue (see text).
BCGs bluer than 0.3 mag from the red sequence at their redshift (i.e. below the solid line, in the
``blue'' region) are presented by means of brighter points. We also
highlight them using the asterisks symbol in the panel for high richness clusters, whereas squares highlight
spirals. 
Dashed lines: 1$\sigma$ region around the mean relation found by Koester et al. (2007) for their BCGs (their
Fig 2, see also Blanton \& Roweis, 2007).}
\label{fig4}
\end{figure}
In Fig.~\ref{fig4} we show the \emph{g-r} color as a function of photometric redshift for all BGCs (upper
panel), whereas we focus on those in rich cluster in the lower panel. We note a clear asymmetry in that the number of BCGs 0.3 bluer (brighter points) than our cutoff,
which is roughly given by the relation $g-r \sim 2.86\cdot z +0.35$ (solid line) is larger than that of the
galaxies on the red side of the average sequence ($g-r \sim 2.86\cdot z +0.65$, dashed lines) .

In order to guide
the eye we also plot - bracketed by lighter dashed lines - the 1$\sigma$ region around the mean relation found by Koester et al. (2007) for their BCGs 
(c.f.their Fig 2) which follow more closely the mean red-sequence evolution as a function of redshift. 
%When
%considering the sample, the difference with Koester et al. is striking, in that the lack of a colour selection let us
%consider BCGs with a wide range of g-r colours (for the difference in the cluster
%properties between the catalogue and other works, instead, refer to Szabo et al., 2010).
On the other hand, the bulk of our BCGs in clusters of richness above 50
matches very well the relation for red and dead early type galaxies. A number of blue
outliers are still present though. We further discuss these galaxies below.

\subsubsection{The blue fraction}

\begin{figure}
%\epsscale{.80}
\includegraphics[width=8cm,height=8cm]{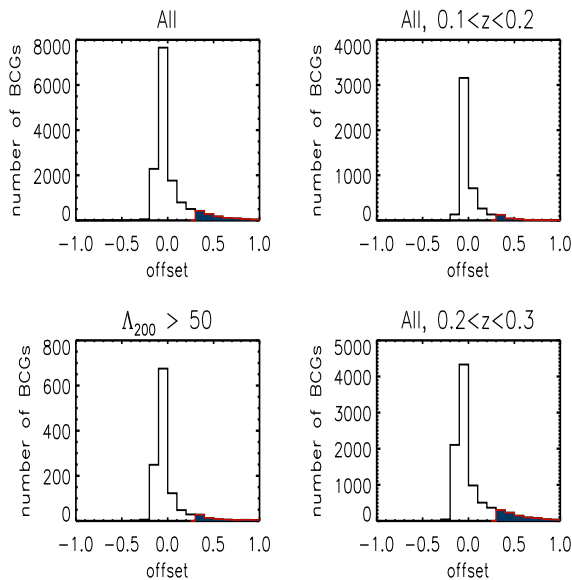}
\caption{Offset (mag) from g-r color-magnitude relation as a function of redshift and offset
distribution in redshift slices. The shaded area emphasizes the tail of blue BCGs.}
\label{fig7}
\end{figure}

We now quantify
the fraction of blue galaxies in terms of redshift, richness and dominance.
In Fig.~\ref{fig7} we show the distribution in the offset in magnitudes from the average g-r 
in different redshift slices. Whilst the large majority of BCGs display a negligible offset
and follow the same trend of the luminous red galaxies studied by Blanton \& Roweis (2007), the
histograms show a clear tail on the blue side (positive \emph{offset} in our figures). As expected from Fig.~\ref{fig4}, 
there is a clear asymmetry in the number of BCGs bluer than the 
red-sequence, with respect to those redder that we showed in Fig.~\ref{fig4}. Note, however,
that the curves become more symmetric and the relevance of the blue BCGs diminishes if we take into
account only rich clusters and as we move to lower redshift.
In particular, we find that the overall fraction of 1st ranked galaxies bluer (redder) than 0.3 mag is 8.6\% (0.2\%). 
Such a fraction does not change if we restrict ourselves to clusters with richness above 30,
and decreases to the 6\% above a richness of 50.
In terms of redshift bins, the blue fraction goes from 5\% in the redshift range 0.1-0.2 to 10\%
in the redshift bin 0.2-0.3. 
%The blue fraction of 1st ranked galaxies exhibits a similar increase (from 10\% to 18\%).
Thus blue BCGs tend to populate poor clusters and their fraction slightly increases with redshift.
Note that at higher redshifts both the scatter in the g-r colour and the its error tend to increase.

The overall blue fraction is lower than the 25\% of BCGs with blue optical cores
found by Bildfell et al. (2008), because in not all cases the spatial extent
of the blue core is large enough to make the entire galaxy blue \footnote{
There are remarkable cases (e.g. Abell 1835) where the BCG can be 0.5-1 mag below
the red-sequence}.
The fraction of 1st ranked galaxies bluer than 0.5 mag is 2.8\%. 
%Therefore, BCG
%catalogues that use a colour-based selection are likely to miss up to 7\% of the BCGs.
%At the same time, sample like the LRG taken as representative of a population
%of massive early-type galaxies may miss up to the 15\% of the actual population.

A common colour-cut used by discriminate between early type morphologies (which
BCGs belong to) from later type at u-r$>$2.22 has been derived by Strateva et al. (2001) on an earlier
release of the SDSS. This requirement is satisfied by the 95\% of our BCGs
in clusters with richness above 30, in agreement with Strateva et al. (2001), who found 
that 97.6\% of their (spectroscopically
classified) early type galaxies were above the u-r=2.2 threshold. 
On the other hand, only the 13\% of those that are blue according to our classification
make this threshold. 
In agreement with us, Choi et al. (2007), found that about 10\% of early type galaxies with
evidences of a blue optical core in a volume limited sample of SDSS has a u-r colour bluer than 2.22. 
Most of these galaxies live in low-density environment as we found.
Unfortunately a cross-match with their catalogue is impossible
because it does not overlap in redshift with ours, being theirs
at redshift below 0.1.

\begin{figure}
%%\epsscale{.80}
\includegraphics[width=8cm,height=8cm]{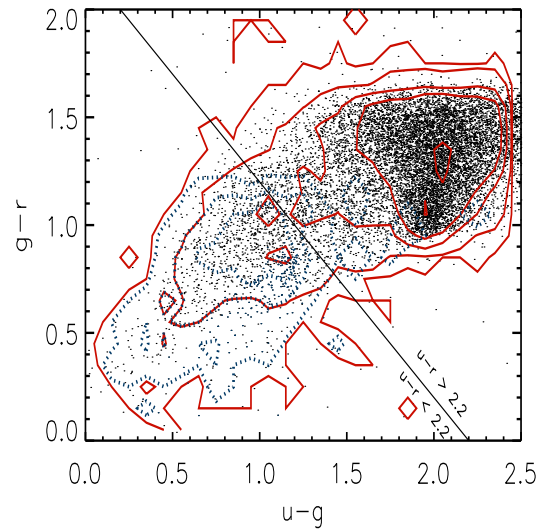}
\caption{Distribution of galaxies in the (g-r) - (u-g) plane. The solid diagonal line
divides the plane at $u-r=$ 2.22: the upper right portion is the typical locus of LRG.
The solid contours pertain to the entire sample of BCGs.
The blue BCGs are presented by the small points and highlighted by dotted contours.}
\label{fig9}
\end{figure}

The presence of \emph{g-r blue} BCGs can be inferred by using other colours.
For instance, galaxies blue in g-r have also quite blue u-g colours, similar
to those of star forming galaxies. This is shown in Fig.~\ref{fig9}.
In this plot the solid diagonal line
divides the plane at $u-r=$ 2.2: the upper right portion is the typical locus of LRG.
The solid contours, that pertain to all the BCGs, peak in this region, 
presenting, however, a large scatter. One natural reason for this is the presence
of the blue BCGs (note secondary peaks at g-r$\sim$u-r$\sim$1 that coincide with the peak of the blue BCG distribution
(dotted contours). Other reasons might be traced back to the fact that there are observational
errors (not taken into
account in this simple analysis). Moreover galaxies are made by mixtures
of stars of different ages and metallicities. This creates a scatter in the colours
of otherwise similar galaxies (for instance galaxies with the same mass and about the same age, see e.g. the Maraston
et al., 2009, but for the LRG case).
Moreover, galaxies are still likely to have a spread in mean ages (namely, it is very unlikely that they have formed
simultaneously at, say, z=3).
%Also, they have a range in, e.g., r-band magnitude, and they obey to 
%colour-magnitude relations, therefore broadly speaking they will be redder (bluer) than the theoretical curve if
%their luminosity is higher (lower) than the luminosity of the model galaxy plotted; moreover, the colour-magnitude relation itself has an intrinsic scatter. 
Blue BCGs cluster in a completely different region of this diagram.
A colour-cut at u-r=2.2 would get rid of most of such galaxies, thus biasing the study of the BCG
population as a whole.
Other colours, such as the i-r, are less useful, since the regions occupied by blue and red galaxies 
tend to significantly overlap. 
%Therefore, it is not surprising that we would find results
%in agreement with Strateva et al. (2001) if we used the i-r colours instead of the u-r.


\subsubsection{The blue fraction in galaxies with spectra}

The fraction of blue BCGs in the whole sample
goes down to 3.7\% when considering only the galaxies that have a spectroscopic
redshift. In particular, the fraction of blue BCGs in cluster with richness above 50
that have a spectroscopic redshift decreases to the 1\%.
In this latter case, all the galaxies with $g-r < 0.7$ would disappear from the lower
panel in Fig.~\ref{fig4} and their fraction would be strongly suppressed in the upper
panel.
We cannot exclude that spectra were preferentially taken for galaxies sitting on the red-sequence, hence
artificially removing blue galaxies, since we have spectroscopic redshifts for just
one third (5642/14344) of the BCGs in the range studied.
%Hence the numbers given above should be regarded as a lower limit of the blue fraction.
However, a close inspection to galaxies in rich clusters tells us that, while
the difference between spectroscopic and photometric redshift tend to be
very small ($<0.02$) in red galaxies, it is not uncommon to find ``blue''
galaxies whose spectroscopic redshift is 0.1-0.2 lower than the photometric estimate.
Given the quite steep variation
of (g-r) with redshift (c.f. Fig.~\ref{fig4}), such a galaxy might be
perfectly on the red-sequence when the correct (spectroscopic) redshift
is used.
Therefore, even considering the \emph{overall} accuracy of state-of-the-art photometric
redshifts estimates, we cannot exclude that a fraction of the blue
BCGs are so because of the \emph{specific} error in their photometric redshift.

Therefore, a conservative final estimate of the blue fraction of BCGs in the
redshift range 0.1-0.3, at richnesses above 50, is between 1 and 6\%.
Such estimate is in agreement with the results by Edwards et al. (2007), who
selected BCGs by means of K-band magnitudes and found that $\sim$3\% of those
had bluer (u-r) colour than expected.

\subsubsection{The blue fraction at higher redshift}

Despite the focus on the present paper is on galaxies in the redshift
range $0.1-0.3$, the parent BCG catalogue has a wealth of data for higher
redshift objects. In this section we briefly quantify the blue fraction
in such a regime. The reader should note that
the colour selection at $z\sim 0.3$ is somehow more complicated. In the first place,
the completeness of the catalogue quickly decreases at $z > 0.4$ , therefore we cannot properly assess the fraction
of blue BCGs. Furthermore, the the g-r -- redshift curve
flattens out already at $z \sim0.35$. Both intrinsic spread and errors in the colours,
as well as the possible presence of evolutionary effects
on the color-magnitude relation, enhance the fraction of blue BCGs at higher redshift.
However, taken at face value, our results in the redshift range $0.3-0.4$\footnote{By adding
galaxies in this redshift range, the total
number of BCGs would nearly double. } 
would indicate that the overall blue fraction increases to 18\% (14\% for clusters with $\Lambda_{200} >50$).
A more conservative estimate can be obtained 
if we consider the fraction of galaxies $\sim 5\sigma$ below the mean g-r at z$\sim$0.4
as the blue ones, namely 0.5 mag instead of 0.3. In such a case the blue
fraction is 11\% (8.5\% clusters with $\Lambda_{200} >50$). These values
are very close to those attained in the 0.1-0.3 redshift range.
A similar increase in the fractions of BCGs with optically blue cores and
emission lines has been observed by Wang et al. (2010) and Crawford et al. (1999), respectively.
As we will see in the following section, our results confirm and extend
the correlation between BCG colour and cluster mass (e.g. More et al., 2010, Loh et al., 2010)
at redshifts z$>$0.1.
%Similar trends and conclusions can be inferred from the g-r Vs r-band magnitude diagrams.

\subsection{Morphologies}

Until this point we did not take into account the morphology of the BCGs.
A more detailed study on the morphologies
in the Szabo et al. catalogue is underway (MacKenzie et al., in prep).
It will tell us whether the blue BCGs are truly early type
galaxies or if the sample is contaminated by late type (i.e. bluer star forming) galaxies. These may be identified
as BCGs if sufficiently bright members of clusters lack, e.g., a well defined red sequence
or a prominent early type central galaxy.
In particular, for application to galaxy formation studies, it can be worthwhile investigating
if the blue fraction can be associated to other galaxy types as opposed
to elliptical ones. For instance, lenticular galaxies are BCGs in known
clusters, while spiral galaxies may dominate groups\footnote{It is worth reminding the reader that relation
between $\Lambda_{200}$ and the cluster mass has a substantial scatter below $\Lambda_{200} = 50$ (Fig. 6 in Dong et al., 2008), 
such that  $\sim 2\cdot 10^{14} h^{-1} M_{\odot}$ poor clusters and $\sim 5\cdot 10^{13} h^{-1} M_{\odot}$
groups can be assigned the same richness.}. 
Finally the blue colours may reflect
the star formation in interacting systems.
Here we mention some preliminary results: in the richest ($\Lambda_{200} > 100$) clusters of our sample,
93\% of the galaxies are single or interacting ellipticals, the rest being ``uncertain''
according to a visual morphological classification.
In particular, the interacting ellipticals alone amount to 34\%, similar
to what found for $z\sim0$ massive ellipticals by Kannappan et al. (2009).
According to Kannappan et al. (2009), these galaxies populate low-to-moderate
density environments. The distributions of these interacting early type galaxies
around the average $g-r$ colour at a given redshift and in absolute magnitude is very similar to 
the one that single ellipticals exhibit.
For the above reasons, we suggest a cut at $\Lambda_{200} > 100$ for studies
focussing on red \emph{early-type} BCGs in massive galaxy clusters.
A visual/spectral classification of the blue BCGs in $\Lambda_{200} > 50$ clusters of our sample,
tells us that the percentage of ellipticals decreases to 70\%.
5\% of these blue galaxies are indeed spirals typically associated to
$\Lambda_{200} \sim 50$ clusters (squares in Fig.~\ref{fig4}).
Being associated to poorer systems, these spirals are on average fainter.
Therefore we suggest a cut at $M_r < - 22.5$ if one wants to use a sample almost made by BCGs that are
early-type galaxies.
The morphology-richness trend seems to follow the increase in the blue fraction at lower richness that we discussed above.
It is worth noting that similar findings are also reported by other studies of the correlation between BCG properties and cluster
mass. For instance, More et al. (2010), using the kinematics of satellite galaxies to infer the halo mass,
found that, at a given galaxy luminosity, red central galaxies tend to occupy more massive
haloes than the blue ones. Similar results are reported by Loh et al. (2010) by means
of the two-point correlation function as well as the NUV-r colours (see below).
Hence, they are able to verify that not only red galaxies but also objects in transition
between the blue and the red sequence, preferentially reside in more massive haloes.
These independent results have been obtained below $z \sim 0.1$, therefore we can confirm the reported trends
and extend their validity out to $z=0.3$.

%We also found two QSOs that contaminate the sample with very extreme ($g-r \sim 0$) colours
%and some galaxies that seems quite isolated.
%Therefore, 
Finally, we cannot exclude that some objects in the entire sample 
are background/foreground field object with peculiar colours.

\section{The BCGs as the three brightest galaxies in a cluster}
In this section we consider the three richest galaxies
as the potential BCGs of a given clusters, because in know
local clusters the brightest is not always the dominant one. Moreover,
massive clusters often show signs of on-going mergers
and clear substructures, each one with its own BCGs.
Finally the importance of comparing the properties
of the brightest galaxy to the second and the third brightest
ones is related to potential applications of the catalogue in
order to constrain galaxy formation models.

\begin{figure}
%%\epsscale{.80}
\includegraphics[width=8cm,height=8cm]{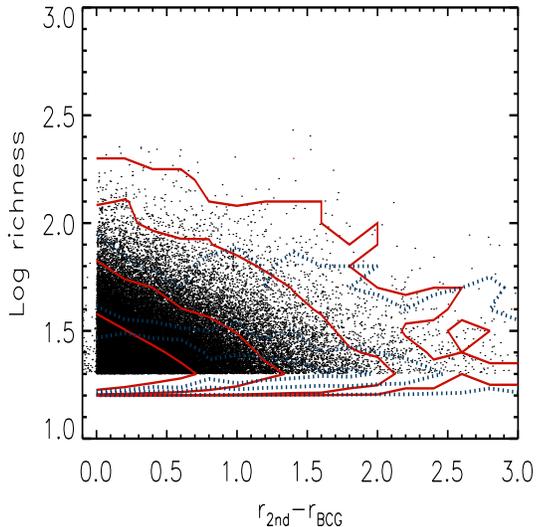}
\caption{Distribution of galaxies in the cluster richness versus 1st ranked BCG \emph{dominance} plane.}
\label{fig3}
\end{figure}

\begin{figure}
%%\epsscale{.80}
\includegraphics[width=8cm,height=8cm]{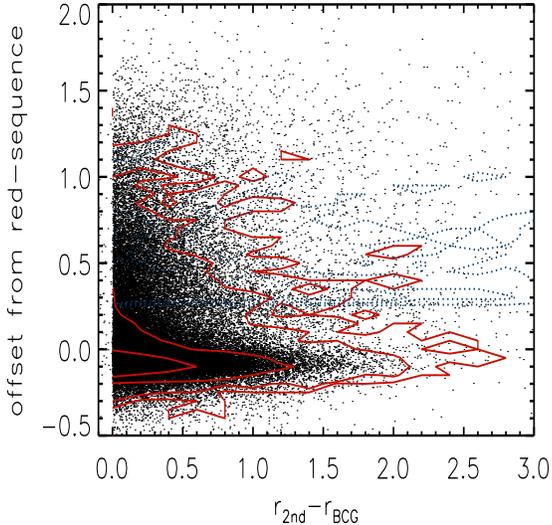}
\caption{Distribution of galaxies in the offset from the red-sequence versus BCG \emph{dominance} plane.
The blue BCGs are presented with dotted contours. The numbers give the actual number of galaxies within that contour.}
\label{fig5}
\end{figure}

\subsection{The role of the \it dominance}

We first turn our attention to the role of the \emph{dominance} of the BCG. We express it in terms of the difference in r-band
apparent magnitude $r_{2nd}-r_{1st}$ between the 2nd and the 1st ranked galaxies.
The average dominance in the redshift range 0.1-0.2 is 0.5 mag in the r-band, a value within
the range given by Loh \& Strauss (2006) \footnote{Note that a direct comparison cannot be made, since
they  focused
only on the LRGs}.
Also our 1$\sigma$ dispersion and the mild decrease of the dominance with redshift are in agreement with 
their findings. 
In Fig.~\ref{fig3} we show the distribution of galaxies in the cluster richness - dominance
plane. 
As expected from other studies (e.g. Loh \& Strauss, 2006), poorer
systems exhibit a larger average dominance than richer clusters.
In the former systems the BCG, albeit less luminous, contributes a larger
fraction of the total light than in the latter. In rich clusters, instead,
the fraction of the total light coming from the BCG is much smaller and
such a difference can tell us something on the path that lead to the formation
of the BCG (e.g. Lin \& Mohr, 2004). In particular,
the dominance as a function of richness, as well as the luminosity
of the brightest and 2nd brightest galaxies as a function of the dominance
are very promising tools to constrain galaxy formation models and to discriminate
among apparently similar recipes for the BCG star formation history (Smith et al., 2010).

Finally, in Fig.~\ref{fig5} we investigate the effect of the dominance on the 
1st ranked galaxy \emph{blueness}, quantified in terms of the offset in magnitude
from the red sequence at the galaxy redshift.
No significant trends are found, except for the fact that cluster
with low dominance are much more abundant than clusters where $r_{2nd}-r_{1st}$ exceeds 1 .

\subsection{The blue fraction}
As far as the blue fractions under such a broader BCG definition are concerned,
we find that it is 12\% overall, 9.8\% in clusters with $\Lambda_{200} >30$
and 8.2\% in clusters with $\Lambda_{200} >50$. Such
values are $\sim$1.5 times larger than the fractions attained
when considering only the brightest cluster galaxy. This fact is
a consequence of the colour-magnitude relation, namely, the fact that
in a given cluster the brightest early type galaxy is also the reddest.
Therefore, in a given cluster, lower luminosity galaxies are likely to be bluer even
if they have an early-type morphology and no signs of star formation.
However we cannot exclude a slightly higher presence of contaminants
as bright spiral galaxies that are in the process of being accreted
in the cluster.
The trend of the blue fraction with redshift in clusters with $\Lambda_{200} >50$
for such a broader definition of BCGs mirrors 
the one found when considering only one (the brightest) galaxy per cluster.

%\subsection{Centrals vs Brighest: HERE? ON IN THE ``APPLICATIONS'' SECTION? TO BE UPDATED ANYWAY}
%
%A non-negligible fraction {\color{red} Thad, how many? how many in rich clusters?}
%of these BCGs are displaced from their host cluster centre by more than 0.5 $R_{vir}$,
%where $R_{vir}$ is the cluster virial radius (see Szabo et al., 2010). The fact that
%the brightest members of a cluster are not always centrally located is not new (e.g. Skibba et al., 2010
%and references therein),
%but implies that a further characterization is needed if one wants to apply
%our findings to constrain the prediction of galaxy-formation models where often the 
%brightest member is assumed to be the central one. Indeed, this is an
%application of the Szabo et al. catalogue that will be the topic of a forthcoming study.
%As an example, we already discussed in Sec. 2.5 that our comparison with known BCGs
%implies that the dominant galaxy in a cluster may be its second or the third brightest member.

\section{A comparison with maxBCG BCGs}

The presence of blue BCGs is an important feature
of the Szabo et al. cluster catalogue. Hence, it is important
to present a quantitative comparison with a widely used
catalogue as maxBCG (Koester et al., 2007) based on 
a strict colour selection.
{We briefly remind here that our BCG is defined as the brightest
galaxy in the r-band that likely belongs to a cluster, even though
it does not necessarily sit at the cluster centre. On the other hand
the maxBCG algorithm requires a red\footnote{We refer to Koester et al., 2007
for the actual colour cuts; here it suffices to say that the criterion
is such that the galaxy colours are within the dashed lines of Fig. 4.} 
and bright galaxy and
at least other ten red and less luminous galaxies within $\sim$ 1 Mpc
to identify a cluster. The seed galaxy is hence the BCG and
the central galaxy at the same time.
The cross-matching between the two catalogues has been done by searching for
maxBCG BCGs that are also one of three brightest galaxies in one of our
clusters.}
In particular, we find that more than 4300 maxBCG are in the entire Szabo et al.'s BCG sample.
In the majority of the cases, there is a one-to-one correspondence, in the sense that
only one maxBCG BCG belongs to a Szabo et al.'s cluster. We define these
clusters as those that match the maxBCG catalogue. For a more thorough description of the matching
procedure and a complete analysis of the outcome we refer to Szabo et al.'s paper.
%We start such an analysis by presenting
%in the upper panel of Fig.~\ref{fig_max} the colours of our 1st ranked BCGs in clusters in common with maxBCG catalogue. 

\begin{figure}
%\epsscale{.80}
\includegraphics[width=8.cm,height=5cm]{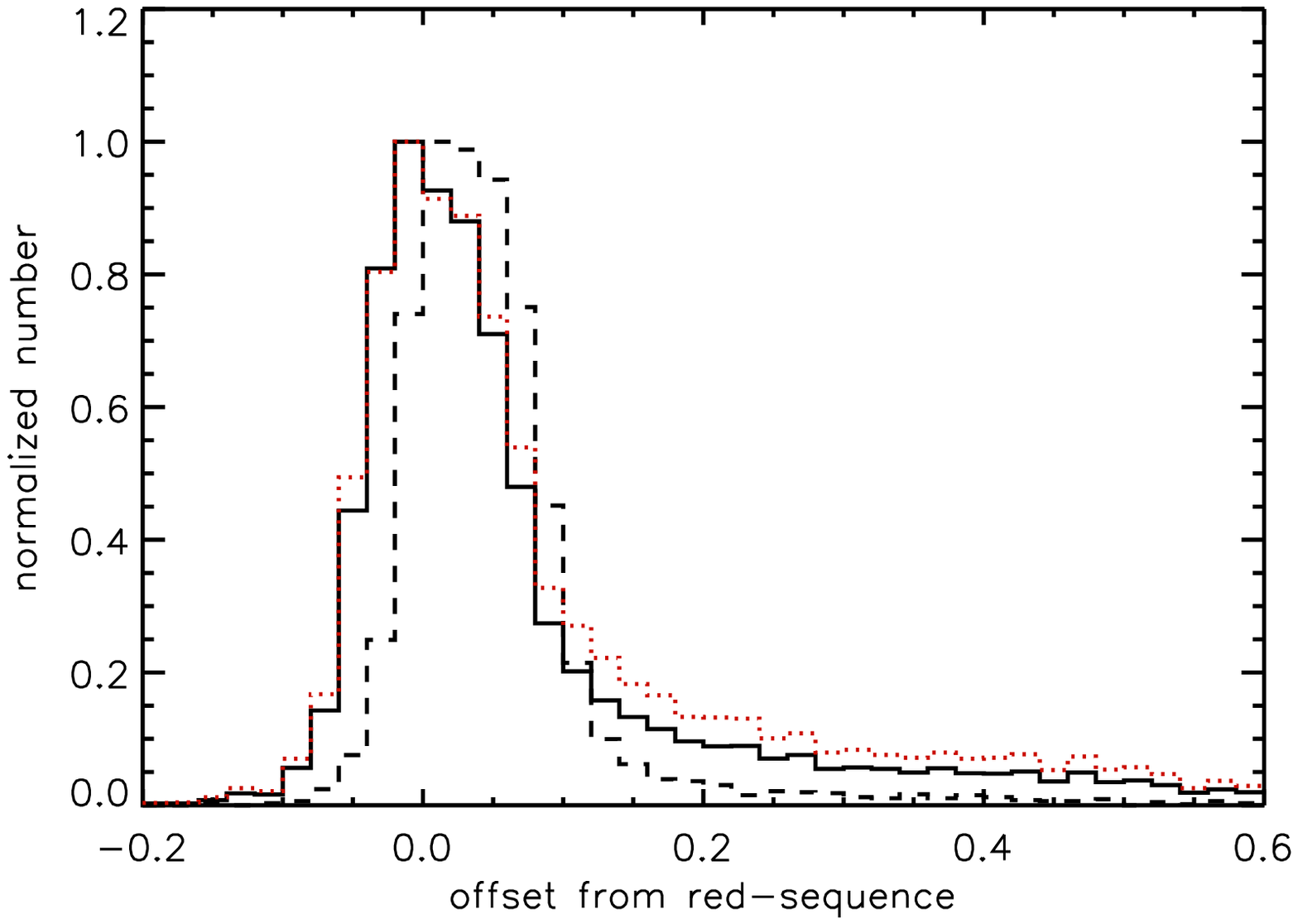}
\includegraphics[width=8.cm,height=5cm]{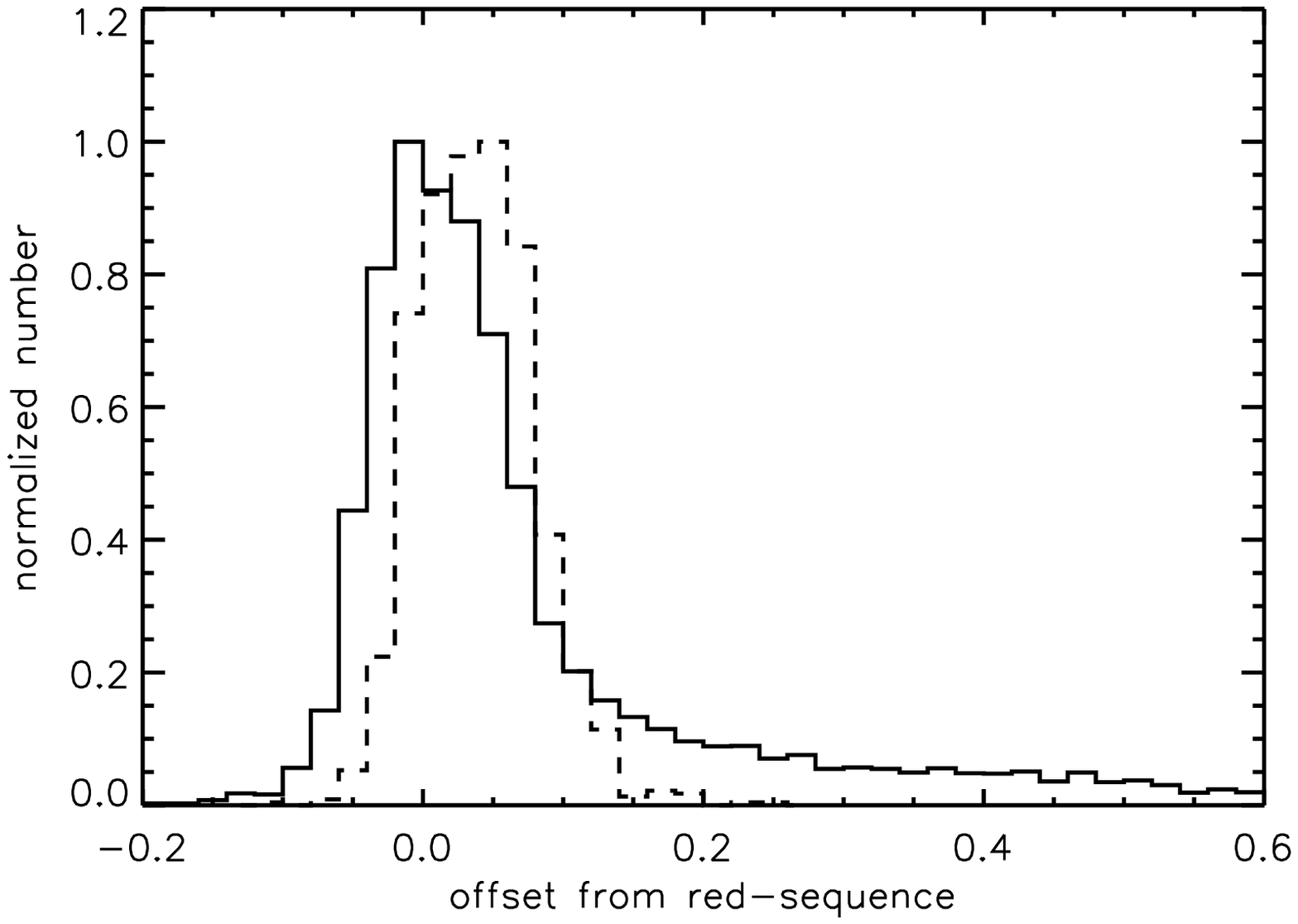}
\caption{\emph{Upper panel:} Offset (mag) from g-r color-magnitude relation
for all our BCGs (solid line) and those that are in clusters in common with maxBCG (dashed - see text) 
and those that are not (dotted line).
%versus the offset from the color-magnitude relation of maxBCG BCGs in clusters that we have in common (dashed line).
\emph{Lower panel:} Offset (mag) from g-r color-magnitude relation
for all our BCGs (solid line, as in the other panel) %and those in clusters not common with maxBCG (solid - see text) 
versus the offset from the color-magnitude relation of maxBCG BCGs that belong to our clusters but that
do not coincide with our BCG (dashed line).}
\label{fig_max}
\end{figure}

In the upper panel of Fig.~\ref{fig_max} we show the distribution in the offset in magnitudes from the average g-r colour
for all our BCGs (solid line) and those that are in clusters in common with maxBCG\footnote{Note that the fact that a fraction of our clusters matches maxBCG clusters
does not imply that our 1st ranked BCG in these clusters always coincides with the BCG of the maxBCG catalogue.}
(dashed line) as well as those that are not in common (dotted line). 
Note that the comparison has been made by taking only Szabo et al's BCG 
in the sky regions where SDSS DR6 and DR5 (on which maxBCG is based) overlap
and in the redshift range 0.1--0.3.
We display normalized
histograms in order to emphasize the tail at positive (blue) offsets in our BCGs.
As expected from the maxBCG algorithm, the colours of maxBCG BCGs that also are BCGs for our clusters do not differ
from those of maxBCG BCGs that belong to our clusters without being the brightest member.
The distribution of our BCGs, instead, shows a more evident tail, because
clusters that do not have a maxBCG counterpart tend to be poor systems (see Szabo et al., 2010),
where the fraction of blue BCG is somehow larger (see above).
The offset distribution for the galaxy that Szabo et al. classify as 1st ranked BCG
in a given cluster and that for the BCGs selected according to the maxBCG method,  
basically coincide but for the tail with blue galaxies.  
Therefore, in clusters that Szabo et al and maxBCG share, the BCG selection
is almost identical. In ~7\% of the cases maxBCG miss the brightest object because is more than 0.3 mag away from
the red-sequence. However, differences are found already at 0.2 mag 
away from the red sequence.
Such a fraction (7\%)  is
close to 6\%, namely is the fraction of blue BCG in clusters with richness $>$ 50 (see above),
and it is a sensible value because most of our matches with maxBCG clusters are in the high richness regime (see Szabo 
et al., 2010).
Since we know that the galaxy classified as BCG by Koester et al. (2007) is still one
of the brightest (but not our 1st ranked) galaxies in our clusters, this means
that maxBCG uses the 2nd
or the 3rd brightest galaxy as seed for their clusters.
Indeed, in the lower panel of Fig.~\ref{fig_max} we plot the offset distribution for 
all our BCGs (solid line, as in the other panel) 
versus the offset from the color-magnitude relation of maxBCG BCGs that belong to our clusters but that
do not coincide with our BCG (dashed line).

In conclusion, if more than 80\% of maxBCG rich clusters have a counterpart in our cluster catalogue (see Szabo et al., 2010), 
we can confirm that they miss a fraction of BCGs bluer than \emph{expected} (namely
0.1 - 0.2 mag off the red-sequence, see Koester et al., 2007).
In other words, the colour criteria used to select maxBCG BCGs may not bias cluster catalogues but certainly affect BCG catalogues
to the 10\% level.
A caveat is that, in order to make the comparison on a the same ``frame'', we used
colours and `` model'' magnitudes for maxBCG BCG as provided by the SDSS DR6. Since
the original maxBCG catalogue has been built from the SDSS DR5, used ``cmodel''
magnitudes and different photometric redshift estimator (and hence
different k-corrections) the exercise presented here does not provide a colour
characterization of maxBCG BCGs in a strict sense.

On the basis of the maxBCG BCG colour alone, instead, we cannot explain why a fair number of maxBCG
clusters is not matched by us. As shown by Szabo et al., the differing completeness above 
a given richness and the differing definition richness between the two catalogues
hamper a 1:1 matching. Also, we cannot exclude that we include groups where
the BCGs are spirals.

%\clearpage
\section{Applications of the catalogue}
\label{uso}

After presenting the general characteristics in terms of optical colours of the Szabo et al. BCG catalogue,
we discuss some possible applications.
To this aim, we also augment the available data per each cluster/BCG  by cross-matching 
with catalogues at other wavelengths.
In particular, we briefly present 
some results on %the BCG \emph{dominance} AND THE CENTRALS?; then study 
the UV-optical colours of our BCGs and
the X-ray properties of their host clusters in connection 
to the galactic \emph{blueness}. In these cases,
since we are matching our sample with existing databases
that are neither complete nor homogeneous, and whose sky coverage
only partially overlaps with ours, we will not yield a complete catalogue.
The following examples cannot, therefore, be taken as a characterization of our
entire catalogue. Hence, in this paper we use them just to highlight the potential
application of our BCG sample in spectral regimes other than the optical.

\subsection{BCGs in the UV}

%Recent work has shown that, at a finer degree of detail, elliptical galaxies are not exactly 
%a class of \emph{red and dead} objects. These results has been derived by comparing the spread in the 
%$NUV-r$ ($\sim$ 6 mag, Fig.~\ref{fig10}) with that in the $g-r$
%color-magnitude relation. 
In contrast to the optical spectral range, the UV is highly sensitive to even
small fractions of young stars (younger than about a Gyr), making
it an excellent probe of the low-level recent star formation (RSF)
that is expected in elliptical galaxies (i.e. star formation
within the last Gyr, contributing up to a few percent of the
stellar mass of the galaxy, e.g. Kaviraj et al., 2007). 
%For instance, using GALEX (UV) and SDSS (optical) photometry, Kaviraj et al.
%(2007a) have shown that in nearby ($0<z<0.11$), massive
%($M(r)<-21$) elliptical population, \emph{at least} 30\% of this
%sample show \emph{unambiguous} signatures of recent star formation.
%In particular, luminous
%ellipticals may form up to 10-15 \% of their stellar mass after
%$z\sim1$, although for the bulk of the population the typical mass
%fractions are much smaller and around a few percent.
As explained earlier, we expect this to happen also
in BCGs, being them the most massive early type galaxies,
since a fraction of them has blue cores (Bildfell et al, 2008, Rafferty et al., 2008)
and shows emission lines (Crawford et al., 1999, Edwards et al., 2007, 2010).

Unfortunately the SDSS database does not provide us with colour gradients to understand if a BCG
in our sample has a blue core due to recent star formation. Nor can the RSF
always lead to blue galaxies in terms of their g-r colour.
Therefore in this section we will make use of the
results put forward by Pipino et al. (2009a, see also Wang et al., 2010).
%In that study, the aim was to understand whether blue cores are
%unambiguously linked to a UV-enhancement, which would indicate
%that the excess UV is driven by recent star formation. In order to
%perform that analysis, Pipino et al. (2009a) cross-matched the 48 BCGs in the Bildfell
%et al. catalog with archival data from the GALEX GR3.
They found that every BCG which has a blue
UV-optical colour also shows a blue-core in its optical colour
profile. Conversely, BCGs that lack blue cores and show monotonic
colour gradients consistent with a decrease in metallicity with
radius typical of old elliptical galaxies are red in the UV. Pipino et al. (2009a)
interpreted this as evidence that the UV enhancement in the blue
BCGs is driven by \emph{recent star formation} and not from old
evolved stellar populations such as horizontal branch stars.
The recent star formation in the blue BCGs typically has an age less than 0.5
Gyrs and contributes mass fractions of less than a percent.
In a sense, the fraction of blue BCG derived in the previous sections
is a lower limit on the RSF in BCGs (or an estimate of the fraction
of BCGs with intense enough RSF to have their optical colours
offset from the red-sequence). The fraction derived in this
section will be, instead, related to the percentage of BCGs
experiencing RSF at a very low level.

We cross-matched the entire BCG sample with publicly
available $UV$ photometry from the GR4 and GR5 data release of the
GALEX mission (Martin et al. 2005) and we find a counterpart
for our entire sample of BCGs in roughly one third of the cases.
In particular, we retrieved data by means of the cross-matched Galex GR4+GR5 -- SDSS catalogue
available in the Galex archive. The positional matching,
performed within 5", returns nearly 5000 objects.
 { The remainder of the galaxies are
not observed partly because the Galex sky coverage only partially overlaps with the SDSS
footprint and because most of the overlap area has been only observed with
All sky Imaging Survey (AIS - short exposure times) rather than with the deeper surveys (Martin et al., 2005).} 
We note that, in general, galaxies harboring recent star formation are more likely
to be detected than non recent star formation-galaxies (since their UV flux will be
higher), especially at higher redshift. Thus at any redshift the
chances of detecting a UV blue BCG is higher than its UV red
counterparts.

%GALEX provides two $UV$
%filters: the far-ultraviolet ($FUV$), centered at $\sim1530\AA$ and
%the near-ultraviolet ($NUV$), centered at $\sim2310\AA$. Note that,
%since our sample is at intermediate redshifts
%($z>0.1$) we only use the NUV filter in this study, since the NUV
%filter traces the spectrum between rest-frame NUV and FUV. 

\begin{figure}
%\epsscale{.80}
\includegraphics[width=8cm,height=8cm]{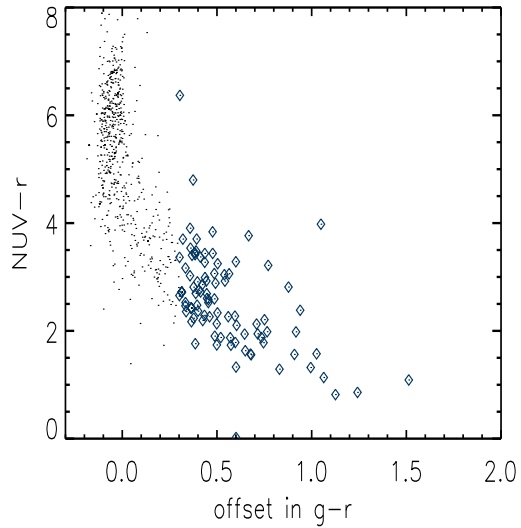}
\caption{NUV-r colour versus offset in magnitudes from the g-r red-sequence (observed frame).
Points: 1st ranked BCGs. Squares: blue BCGs. Note: to improve the quality, only a limited number of galaxies in the region with g-r$\sim$0 and NUV-r$\sim$6 is plotted.}
\label{fig11}
\end{figure}

Fig.~\ref{fig11} shows that optically blue BCGs systematically have
a UV excess (i.e. NUV-r $\sim$ 4\footnote{It is worth reminding that NUV-r $<$5.3 is
a conservative limit that discriminates between UV excess caused
by recent star formation, and that caused by old stars (Yi et al., 2005).}). 
In particular, we find that while only the 14.7\% of BCGs
(that have a GALEX counterpart) in cluster with richness higher than 30 exhibit a NUV-r colour below 3,
this fraction rises up to 71\% when considering only the optically blue ones.
Very similar fractions (12.5\% and 72\%, respectively) are obtained
considering BCGs in clusters whose richness exceeds 50.
Note that in the NUV-r vs r color-magnitude diagram the dispersion is much larger than
typical optical colours colour-magnitude relation; therefore 
the threshold at NUV-r=3 used here emphasizes the 
galaxies with the bluest UV-optical colour.

%\begin{figure}
%%\epsscale{.80}
%\includegraphics[width=8cm,height=8cm]{fig10.eps}
%\caption{NUV-r colour versus r-band magnitude (observed frame). Points: all BCGs.
%Squares: blue BCGs in the entire sample.  Here
%the red-sequence roughly corresponds to galaxies with NUV-r$\sim$7.}
%\label{fig10}
%\end{figure}

\begin{figure}
%\epsscale{.80}
\includegraphics[width=8cm,height=8cm]{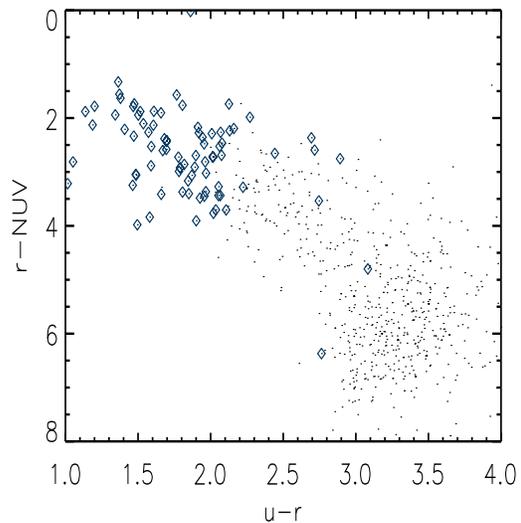}
\caption{NUV-r colour versus u-r colour. Symbols as in Fig.~\ref{fig11}.}
\label{fig12}
\end{figure}

{Choi et al. (2007, c.f. their Fig. 6) show that colour gradients correlate with
u-r colour for SDSS early type. They find that $\sim$ 10\% of morphologically classified
ellipticals have positive colour gradients and u-r $<$ 2.2.
In particular, they find  gradients that are positive (blue cores) at
u-r $<$ 2.2 and becoming more and more negative (red and dead ellipticals) as the
u-r colours gets redder. This is exactly what we find and show in our
Fig.~\ref{fig12} if we assume that there is a one-to-one correspondence
between the optical blue core and the UV excess (Pipino et al., 2009a).
This correspondence is suggestive, however a more detailed work is needed
to establish it on solid grounds. }
%As shown
%by Pipino et al. (2009a) there is a one-to-one correlation between UV-optical \emph{blueness} and the presence
%of a blue core. In most of the cases, then, the blue core makes the entire galaxy bluer
%than the red-sequence.
For instance, one should perform a deeper
analysis that includes a visual morphological classification, 
an inspection of all the matches to ensure
that we do not include UV light coming from close companions, foreground and background
(including lensed) objects. This is left to a forthcoming study (MacKenzie et al., in prep.).

\subsection{BCGs in the X-ray}

Bildfell et al. (2008) found
that the presence of optical blue cores in 25\% of its BCG sample is directly linked to the
X-ray excess of the host clusters. Moreover the position of these BCGs
coincides with the peak in X-ray emission. 
Their interpretation is that the recent
star formation in BCGs is associated with the balance between heating and cooling in the
ICM in the sense that the clusters that are actively cooling are forming
stars in their BCGs.
The aim of this section is to confirm the results by Bildfell et al. (2008) 
and show how our BCG catalogue can be successfully exploited for studies
of X-ray clusters and intergalactic medium properties.
For this purpose, we use the entire catalogue of (1st ranked) BCGs from Szabo et al., without
any redshift limit.

We make use of the publicly available data made possible by the ACCEPT project (Cavagnolo et al. 2009).
The catalog comprises 239 galaxy clusters with accurate temperature, density, entropy and pressure
profiles reduced in a homogeneous way from public Chandra data. The catalog covers
the temperature range 1-20 keV with redshifts ranging from 0.05 to 0.89.
We refer to Cavagnolo et al. (2009,
and references therein) for details on the data reduction and further catalog specifics.
Here we note that their catalog is neither flux limited, nor volume limited. 
{The matching procedure is similar to what done in Sec.~\ref{test}, however a more detailed scrutiny
is required. In the first place, in order to maximize the number of matches we do not
limit the analysis to BCGs in the redshift range 0.1-0.3. 
%Secondly, since Bildfell et al. (2008)
%carefully always chose the galaxy among the cluster brightest members which is also
%the closest to the X-ray centroid, we extend the cross-matching procedure to the top three
%brightest galaxies in each AMF cluster. 
We discard matches that have a difference
in redshift larger than 0.03 if the galaxy spectroscopic redshift is available, 0.1 otherwise\footnote{This
is required because bluer galaxies sometimes have an overestimated photometric redshift.}.
In this latter case, however, we further check that the galaxies are associated
to the clusters. In particular, 
we visually inspect the SDSS images in order to study the position of the galaxy in relation to
the literature position of the X-ray cluster. In practice,
the match is rejected when a clear visual over-density of galaxies is found around the
literature X-ray cluster position \emph{but} the alleged BCG is instead isolated
and more distant. }

\begin{figure}
%\epsscale{.80}
\includegraphics[width=8cm,height=8cm]{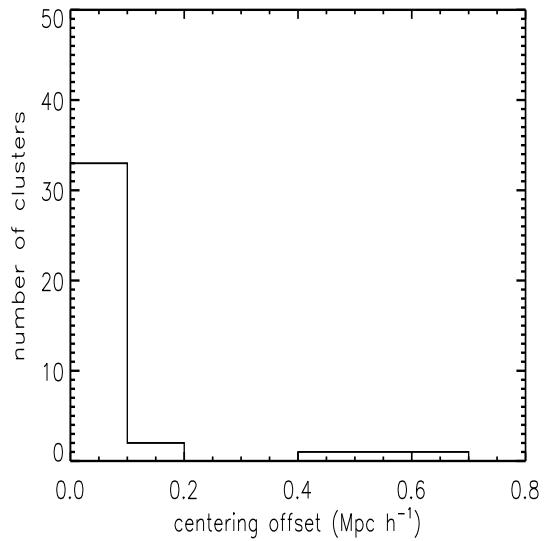}
\caption{Positional offset between the BCG and the cluster X-ray centre for clusters
in common with the ACCEPT sample.}
\label{fig13}
\end{figure}

Such a procedure returns 38 matches
of which 35 within 200 kpc from the cluster centres (Fig.~\ref{fig13}) . This is a well-known properties of BCGs, i.e.
they are typically located within few arc seconds from the cluster centres.
%If we limit ourselves at the redshift range 0.1-0.3
%the number of available clusters decreases to 20 . In the following we focus on these.

\begin{figure}
   \includegraphics[width=8cm,height=8cm]{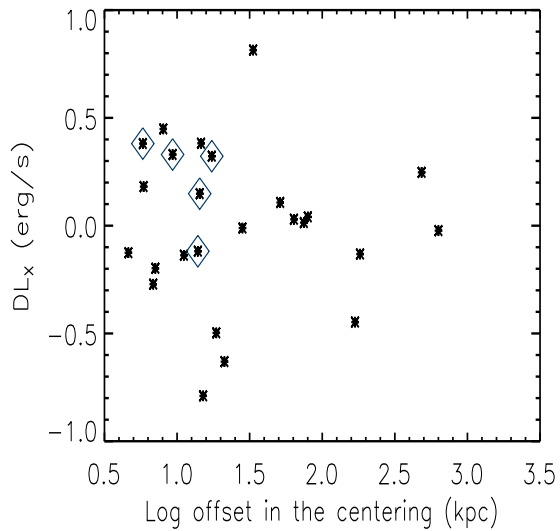} 
   \caption{X-ray luminosity excess vs. BCG/X-ray positional offset for clusters that we have in common with the ACCEPT sample.
NUV blue (i.e. NUV-r $<$ 5) BCGs are identified with an asterisk symbol surrounded by a diamond. 
There is an obvious tendency for star forming BCGs to lie closest to their host 
cluster's X-ray peak while the normal red non star forming BCGs are the furthest. Moreover, blue BCGs are in clusters
that lie above the mean $L_X$-$T_X$ relation.}
   \label{fig14}
\end{figure}

\begin{figure}
%%\epsscale{.80}
\includegraphics[width=8cm,height=8cm]{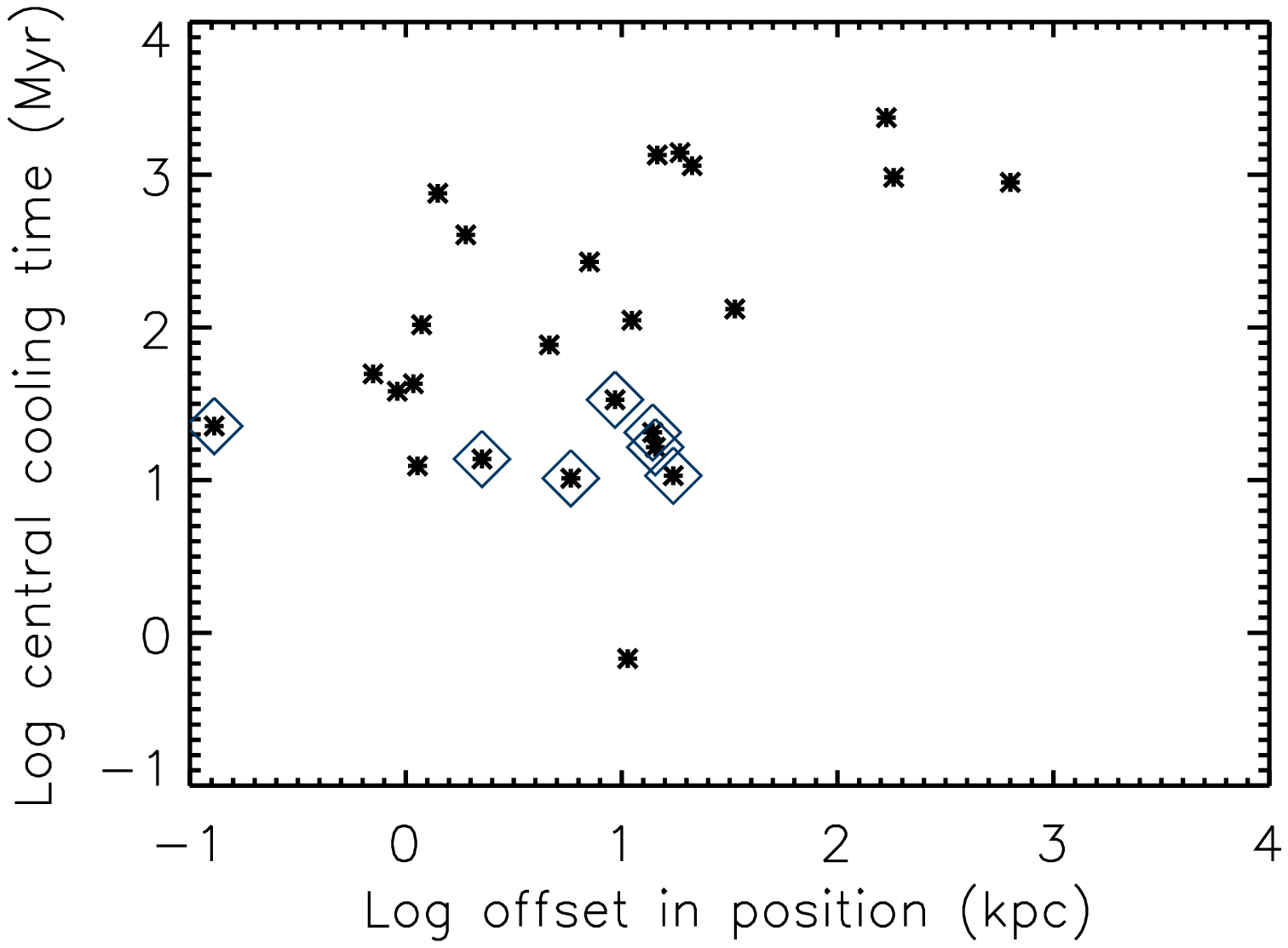}
\includegraphics[width=8cm,height=8cm]{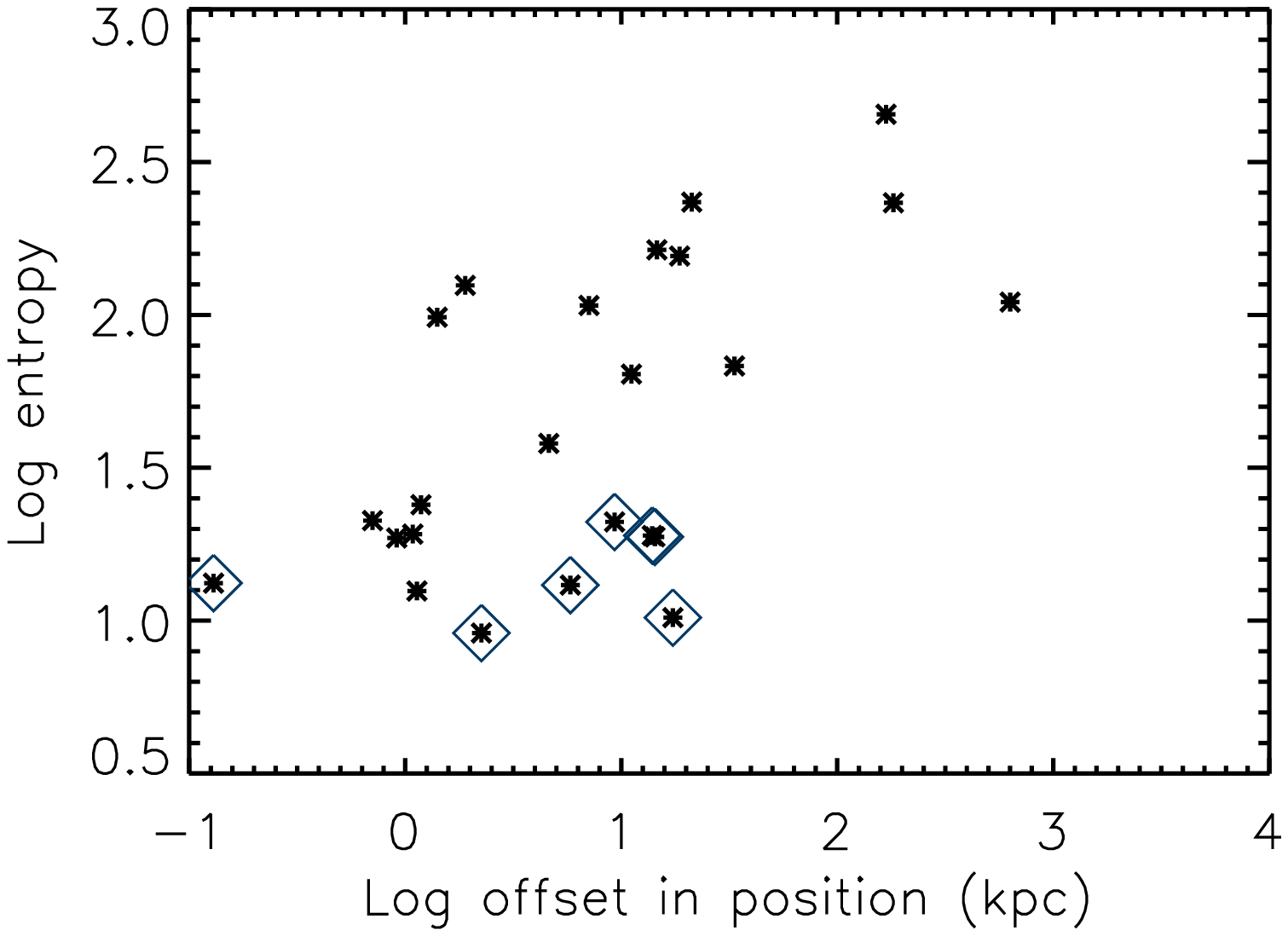}
\caption{Cluster cooling time (upper panel) and entropy (lower panel) versus BCG's offset from the the cluster centre. 
Symbols as in the previous figure. 
\emph{NUV blue} (i.e. NUV-r $<$ 5) BCGs are hosted by cluster with short cooling times and are at low distances from the cluster centre.}
\label{fig16}
\end{figure}

\begin{figure}
%\epsscale{.80}
%\includegraphics[width=8cm,height=8cm]{fig15.eps}
\includegraphics[width=8cm,height=8cm]{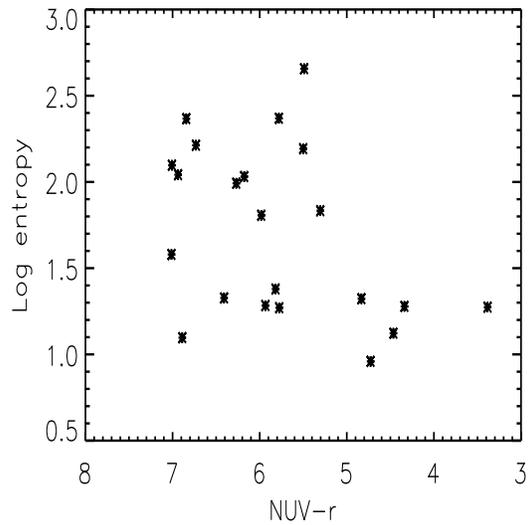}
\caption{Cluster excess entropy versus BCG's NUV-r colour for clusters
in common with ACCEPT. }
\label{fig15}
\end{figure}

We compute the mean $L_X$-$T_X$ relation for the clusters that we match by means of a linear
regression. If we look at the positional offset versus offset in $L_x$ from the mean $L_x$-$T_x$ relation at a given $T_x$ (given
by the quantity $D{L_x}\equiv\Delta log(L_x/E(z)$ $ h^{-2}_{70} erg s^{-1})$) presented
in Fig.~\ref{fig14}, we note that the NUV blue BCGs (asterisks surrounded by a diamond) tend to cluster at positive values of X-ray excess
and very small distances from the cluster centres. This was the diagnostic used by Bildfell et al. (2008) to infer
the link between cooling , galaxy position and galaxy blue core.
Here we are in the position of showing that such a conclusion is supported by the analysis
of the cooling times (Fig.~\ref{fig16}, upper panel) and entropy (Fig.~\ref{fig16}, lower panel)
as provided by Cavagnolo et al. (2009).
Therefore we can confirm Bildfell et al. (2008) results that a fraction of blue
BCGs can be explained by %their proximity with the X-ray 
cluster centres where cooling
flows supply cold gas for star formation. 
%In practice, Fig.~\ref{fig15}, BCGs featuring NUV-r $<$ 5 have entropies below 30 keV$rm /cm^2$. 
That is, BCGs in low entropy clusters may not have enough star formation to
be optically blue, but they are definitely below the NUV-r red-sequence (Fig.~\ref{fig15}).

In a companion paper (Pipino \& Pierpaoli, 2010) we further link the presence
of a cooling flow (and hence the likely presence of a blue BCG) with an enhanced Sunyaev-Zeldovich (Sunyaev \& Zeldovich, 1970) signal in order
to study the bias induced in SZ-based cluster catalogues.
Finally, a more general comparison of our cluster catalogue with nearly 1000 clusters
observed in the X-rays at a lower spatial resolution and available
in the literature is presented in the main catalogue paper (Szabo et al., 2010),
where the correlation between richness and relevant quantities such as
$L_X$ and $T_X$ is made.

\section{Conclusions}
\label{concl}

In this paper we characterize in terms of magnitude and colours a sub-sample of more than 14300 BCGs
drawn from a parent catalogue of more than 220000 galaxies in 69000 clusters
based on the matched filter method (Kepner
et al., 1999, Dong et al., 2008) applied to the SDSS DR6 (Szabo et al., 2010) 

In agreement with previous works, the BCG luminosity is found to a have a redshift evolution
broadly consistent with pure ``aging'' of the galaxies. Richer clusters
tend to have brighter BCGs, however less \emph{dominant} than in poorer
systems.
4-9\% of our BCGs are at least 0.3 mag bluer
in the g-r colour than the red-sequence at their given redshift.
Such a fraction does not change if we restrict ourselves to clusters with richness above 30,
and decreases to the 1-6\% above a richness of 50. In this
case 3\% of them are 0.5 mag below the red-sequence. 
A preliminary morphological study suggests that the increase
in the blue fraction at lower richnesses has a contribution for the increase in the fraction 
of spiral galaxies. Therefore we suggest a cut at richness
of 100 (and $M_r < -22.5$ mag) if one wants to focus only on early-type galaxies. 
In terms of redshift evolution, the overall blue fraction goes from $\sim$5\% in the redshift range 0.1-0.2 to $\sim$10\% in the redshift bin 0.2-0.3. 
The blue fraction seems to increase at higher redshifts, however the scatter in the colours
and the fact that the catalog is no longer complete hamper us from having firm conclusions.
We show that a colour selection based on the g-r red-sequence or on a cut at colour u-r$>2.2$ can lead to missing the majority of such blue BCGs.
Finally, the blue fraction increase by a factor 1.5 at most when the study is extended
to the three brightest galaxies of each cluster.

The fraction of blue BCGs is in broad agreement with previous works (e.g., Crawford et al., 1999, Edwards et al., 2007, Bildfell et al., 2008) which showed that
about one quarter of BGCs show emission lines and optical blue cores associated with
recent star formation; only a smaller fraction has the star formation
extended (in time and space) enough to make their total colour blue.

We also show two applications of the Szabo et al. BCG catalogue.
The first extends the colour analysis to the UV range by cross-matching
our catalogue with publicly available data from Galex GR4 and GR5.
We show a clear correlation between offset from the 
optical red-sequence and the amount of UV-excess.
Pipino et al. (2009a) showed a one-to-one correlation between optical
blue cores created by some residual star formation and UV-excess in ellipticals. Therefore we can
infer that 8\% of our BCGs are offset from the main
sequence because of central recent star formation. The fraction
of BCGs with even lower residual star formation that can be noticed
only by means of UV-optical colours can be as high as 15\%.

{ We compare the colours of our BCGs in clusters in common with the maxBCGs (Koester
et al., 2007) catalogue.
In ~7\% of the cases maxBCG miss the brightest object because is more than 0.3 mag away from
the red-sequence. In this cases, the galaxy classified as 
BCG by Koester et al. (2007) is still one of the brightest (but not our 1st ranked) galaxies in our clusters, this means that maxBCG uses the 2nd
or the 3rd brightest galaxy as seed for their clusters. Such a difference
is not enough to explain the differences between the AMF and the
maxBCG catalogues (see Szabo et al., 2010).}

We cross-match our catalogue with the ACCEPT cluster sample (Cavagnolo et al., 2009),
where accurate temperature, density and entropy profiles of the Intracluster
medium can be found.
We find that blue BCGs tend to be %lie within 100 kpc from the cluster centre
in clusters with low entropy, short cooling times. 
%and typically with
%a $L_X$ above the mean for their temperature. 
That is, the blue light
is presumably associated to gas feeding of recent star formation by cooling flows (Bildfell et al., 2008).

%The facts that we do not exclude blue BCGs and that these galaxies tend
%to lie very close to the X-ray cluster centres, enhance the
%cross-correlation and positional matching of our catalogue with
%published compilations of X-ray selected clusters with respect to other
%works.

%.................................................................................................................

\section*{Acknowledgments}
%The authors thank the referee for his careful reading and his insightful comments.
AP wishes to thank S.Ameglio for many enlightening discussions and L. Edwards, J. Gunn and M. Rich for
useful comments.

AP, TS and EP acknowledge support from NSF grant AST-0649899. 
%AP is also supported by the Italian Space Agency
%through contract ASI-INAF I/016/07/0EP; 
EP is also supported
by NASA grant NNX07AH59G and JPL-Planck subcontract 1290790
and thanks the Aspen Center for Physics.
SMacK acknowledges support from NSF grant NSF-PHY 0850501.

This research has made use of: the SIMBAD database, operated at CDS, Strasbourg, France;
the X-Rays Clusters Database (BAX), which is operated by the Laboratoire d'Astrophysique de Tarbes-Toulouse (LATT),
under contract with the Centre National d'Etudes Spatiales (CNES); the NASA/IPAC Extragalactic Database (NED) which is operated by the Jet Propulsion Laboratory, California Institute of Technology, under contract with the National Aeronautics and Space Administration.
Data presented in this paper were obtained from the Multimission Archive at the Space Telescope Science Institute (MAST). STScI is operated by the Association of Universities for Research in Astronomy, Inc., under NASA contract NAS5-26555. Support for MAST for non-HST data is provided by the NASA Office of Space Science via grant NAG5-7584 and by other grants and contracts.

GALEX (Galaxy Evolution Explorer) is a NASA Small Explorer,
launched in April 2003, developed in cooperation with the Centre
National d'Etudes Spatiales of France and the Korean Ministry of
Science and Technology.

Funding for the SDSS and SDSS-II has been provided by the Alfred
P. Sloan Foundation, the Participating Institutions, the National
Science Foundation, the U.S. Department of Energy, the National
Aeronautics and Space Administration, the Japanese Monbukagakusho,
the Max Planck Society, and the Higher Education Funding Council
for England. The SDSS Web Site is http://www.sdss.org/.

%.................................................................................................................

\end{document}